\documentclass[acmsmall,nonacm]{acmart}

\setcopyright{none}
\authorsaddresses{}

\usepackage{placeins}
\usepackage{algorithm}
\usepackage{algpseudocode}
\AtBeginDocument{%
  }

\begin{document}

\title{Feedback-Driven Rate Control for Learned Video Compression}

\author{Zhiheng Xu}
\email{244512051@csu.edu.cn}
\affiliation{%
  \institution{School of Electronic Information, Central South University}
  \city{City}
  \country{China}
}

\author{Xuerui Ma}
\email{maxuerui@mlslabs.com.cn}
\affiliation{%
  \institution{Malanshan Audio \& Video Laboratory}
  \city{City}
  \country{China}
}

\author{Chunhua Peng}
\authornote{Corresponding author.}
\email{pch@csu.edu.cn}
\affiliation{%
  \institution{School of Electronic Information, Central South University}
  \city{City}
  \country{China}
}

\author{Hao Zhang}
\authornote{Corresponding author.}
\email{hao@csu.edu.cn}
\affiliation{%
  \institution{School of Electronic Information, Central South University}
  \city{City}
  \country{China}
}

\renewcommand{\shortauthors}{Xu et al.}

\begin{abstract}
In recent years, end-to-end learned video compression has achieved significant progress in rate--distortion performance. However, rate control for learned video coding remains relatively underexplored, especially in target bitrate-driven and budget-constrained coding scenarios. Existing methods mainly rely on explicit R--D--$\lambda$ modeling or feed-forward prediction networks to determine coding parameters, but they may lack stable online adjustment capability when video content varies dynamically.

To address this issue, this paper proposes a rate control framework for learned video compression based on closed-loop feedback regulation and rate--distortion optimization. First, we build a single-model multi-rate coding interface on top of a DCVC-style learned video compression framework, enabling the encoder to cover multiple bitrate operating points by adjusting the rate--distortion parameter $\lambda$. On this basis, a log-domain PI/PID closed-loop control algorithm is introduced to adjust the coding parameter $\lambda$ online according to the error between the target bitrate and the bitrate estimated by the entropy model, thereby achieving stable and attainable target bitrate tracking.

Furthermore, to improve the rate--distortion efficiency of frame-level bit allocation under budget constraints, we propose a budget-constrained RD-optimized adjustment controller based on a dual-branch GRU. Taking the base control signal generated by the PI/PID algorithm as the anchor, this module uses budget-state features and causally available coding statistics as inputs to refine the coding parameter $\lambda$, thereby optimizing inter-frame rate--distortion allocation while maintaining overall bitrate stability.

Experimental results on the UVG and HEVC standard test sets show that the proposed PI/PID algorithm achieves stable target bitrate control, with average bitrate errors of approximately 2.88\% and 2.95\% on DCVC and DCVC-TCM, respectively. After introducing the proposed adjustment controller, our method achieves average BD-rate reductions of 5.69\% and 4.49\% on the single-model multi-rate backbones based on DCVC and DCVC-TCM, respectively, while further reducing the average bitrate errors to 2.13\% and 2.24\%. These results demonstrate that the proposed method provides a practical solution for learned video compression that jointly offers controllable bitrate and improved rate--distortion optimization capability.
\end{abstract}



\maketitle

\section{Introduction}
\label{sec:introduction}

Efficient video compression is of great importance for storage, transmission, and downstream vision applications. Traditional video coding standards, such as H.264/AVC and HEVC, have established mature frameworks for prediction, transform, quantization, and entropy coding, together with relatively complete rate control methodologies, including rate--distortion modeling, $\rho$-domain control, $\lambda$-domain control, and PI/PID-based feedback regulation~\cite{wiegand2003overview,sullivan2012overview,ma2005rate,wang2013quadratic,li2014lambda,wong2004pid,shen2009frame,zhou2011pid,meng2016adaptive}. In contrast, although end-to-end learned video compression has achieved remarkable progress in rate--distortion performance~\cite{lu2019dvc,hu2021fvc,li2021deep,sheng2022temporal,li2022hybrid,li2023neural,yang2024learned,tang2024high,sheng2025bi,liao2025ehvc}, its bitrate controllability remains relatively limited.

Most existing learned video compression methods obtain different RD operating points by using fixed $\lambda$ values, or rely on discrete quality levels to cover multiple bitrates, making it difficult to achieve stable bitrate control under a given target bitrate in the same way as traditional encoders. In recent years, some studies have begun to investigate single-model multi-rate modeling and structure-adaptive coding~\cite{lin2021deeply,chen2023sparse,wang2026dual,yang2024adaptive,tang2025uvc}, while others have explored the rate control problem in learned video compression~\cite{li2022rate,zhang2023neural,liao2024rate,zhang2024learned}. Meanwhile, several works have further extended the controllability and adaptability of learned video compression from the perspectives of scene content adaptation, practical real-time coding, and compression control for downstream vision tasks~\cite{xu2024neural,jia2025towards,reich2024deep}. However, explicit feedback-based, online-updating, and deployment-friendly frameworks, similar to those widely used in traditional video coding, have been much less explored.

To this end, we propose a feedback-driven rate control framework for learned video compression. First, we apply and further extend the method of Lin \emph{et al.}~\cite{lin2021deeply} to DCVC by incorporating $\lambda$-conditioned modulation into the context-dependent coding path of DCVC, enabling a single model to cover multiple bitrate operating points and providing a continuously adjustable rate--distortion interface for subsequent control. On this basis, we introduce the PI/PID feedback control paradigm into end-to-end learned video compression, where the entropy-model-estimated bitrate is used as the feedback signal in the log domain to adjust $\lambda$ online on a frame-by-frame basis, thereby achieving stable and attainable bitrate tracking under a given target bitrate constraint. Furthermore, relying solely on the basic closed-loop controller can guarantee bitrate stability, but cannot optimize bitrate allocation according to video content. Therefore, we further propose a budget-constrained RD-optimized adjustment controller. Anchored to the base control signal generated by the PI/PID controller, this module predicts frame-wise $\Delta \log \lambda$ through a dual-branch GRU, so as to further improve inter-frame bit allocation while preserving the stability of the underlying rate control process.

Experimental results show that the proposed method not only achieves low bitrate error on UVG and multiple HEVC standard test sets, but also attains better rate--distortion performance than fixed-$\lambda$ coding, demonstrating the effectiveness of combining explicit feedback control with budget-constrained optimization in learned video compression.

The main contributions of this paper are summarized as follows:
\begin{itemize}
    \item We extend $\lambda$-conditioned modulation to the context-dependent coding path of DCVC, thereby establishing a continuously adjustable control interface.

    \item We introduce the PI/PID closed-loop rate control paradigm into the learned video compression framework to achieve stable online tracking of target bitrates, and adopt a more robust PI configuration in our implementation.

    \item Under budget constraints, we propose a budget-constrained RD-optimized adjustment controller. Anchored to the base control signal, this module predicts frame-wise $\Delta \log \lambda$ adjustments through a dual-branch GRU, thereby further improving the efficiency of inter-frame bit allocation while maintaining the stability of the underlying bitrate control.
\end{itemize}

\section{Related Work}
\label{sec:related_work}

\subsection{Learned Video Compression}

End-to-end learned video compression (LVC) has developed rapidly in recent years. Early works, such as DVC~\cite{lu2019dvc} and FVC~\cite{hu2021fvc}, unified motion estimation, motion compensation, motion information compression, and residual compression into an end-to-end rate--distortion optimization framework, thereby establishing the basic paradigm of neural video compression. Building upon this foundation, subsequent studies have continuously improved compression performance by enhancing spatiotemporal context modeling, motion compensation accuracy, and the expressive capability of entropy models. Representative works include DCVC~\cite{li2021deep}, TCM~\cite{sheng2022temporal}, Hybrid Spatial-Temporal Entropy Modelling~\cite{li2022hybrid}, Diverse Contexts~\cite{li2023neural}, as well as further advances on spatiotemporal priors, adaptive temporal modeling, and unified compression frameworks by Yang \emph{et al.}~\cite{yang2024learned} and Tang \emph{et al.}~\cite{tang2025uvc}. At the same time, recent studies have also introduced modules such as attention mechanisms, feature-level enhancement, and context filtering to further improve compression quality, including High Visual-Fidelity LVC~\cite{li2023high}, the spatiotemporal cross-covariance transformer-based method~\cite{chen2023neural}, FLAVC~\cite{zhang2025flavc}, and the method with in-loop contextual filtering~\cite{wu2025neural}. In addition, existing research has further expanded to bidirectional prediction and hierarchical reference structures, such as Bi-DCVC~\cite{sheng2025bi}, EHVC~\cite{liao2025ehvc}, Learned Bi-Directional Motion Prediction~\cite{shi2022learned}, IBVC~\cite{xu2024ibvc}, and MH-LVC~\cite{phung2025mh}. Overall, existing LVC methods mainly focus on improving rate--distortion performance, while research on achieving online controllable coding under a given target bitrate constraint remains relatively limited.

\subsection{Variable-Rate and Rate Control for Learned Video Compression}

To improve the capability of a single model to cover multiple bitrate operating points, variable-rate modeling has gradually become an important research direction in learned compression. Lin \emph{et al.}~\cite{lin2021deeply} proposed a variable-rate video compression method based on $\lambda$-conditioned modulation, where $\lambda$ is used as a conditional input to modulate internal network features, enabling a single model to cover multiple bitrate operating points. Subsequently, several studies further enhanced the multi-rate capability of a single model from the perspectives of architecture design and feature synchronization, such as the dual-scale transformer-based variable-bitrate synchronization method~\cite{wang2026dual}, as well as methods that improve coding flexibility through prediction-structure adaptation~\cite{yang2024adaptive}.

For rate control in learned video compression, existing studies have begun to explore different forms of rate--distortion modeling and control strategies. For example, Li \emph{et al.}~\cite{li2022rate} proposed a rate control method for LVC by establishing an R-D-$\lambda$ model for frame-level bitrate allocation and adopting a staged parameter update strategy to improve the stability of parameter estimation. Subsequently, Neural Rate Control~\cite{zhang2023neural}, Rate-Quality Based Rate Control~\cite{liao2024rate}, Adaptive Rate Control with RD Prediction~\cite{gu2025adaptive}, and frame-level adaptive rate control based on dynamic neural networks~\cite{zhang2024learned} have advanced learned rate control from the perspectives of rate--distortion prediction, budget modeling, and dynamic network routing, respectively. Meanwhile, some studies have improved coding flexibility through scene-content-adaptive networks~\cite{xu2024neural}, or expanded the adjustable range and improved the rate--distortion performance of a single model through scale adaptation and sparse-to-dense strategies~\cite{chen2023sparse}. In addition, recent works have also paid increasing attention to efficiency and real-time performance in practical deployment scenarios of learned video compression~\cite{jia2025towards}, as well as the impact of compression control on downstream vision tasks~\cite{reich2024deep}. Although these methods have promoted the development of learned rate control, most of them still rely on feed-forward prediction or policy selection, and research on online feedback-based bitrate regulation remains relatively limited.

\subsection{Classical Feedback Control for Video Rate Regulation}

In traditional video coding, rate control has long been one of the core research problems. For standards such as H.264/AVC and HEVC, researchers have developed various rate control methods, including rate--distortion analysis-based approaches~\cite{ma2005rate}, $\rho$-domain control~\cite{wang2013quadratic}, and $\lambda$-domain control~\cite{li2014lambda}. On this basis, feedback control has also been introduced into video bitrate regulation. For example, PI/PID control has been adopted to update coding parameters online according to the error signals generated during the encoding process~\cite{wong2004pid,yu2002new,shen2009frame,zhou2011pid,meng2016adaptive}.

However, the rate control problem in learned video compression (LVC) differs significantly from that in traditional coding. Existing methods mainly rely on explicit rate--distortion modeling or feed-forward prediction mechanisms, while studies on online feedback-based bitrate regulation remain relatively limited. Therefore, introducing feedback control mechanisms into learned video compression frameworks to achieve more stable and controllable bitrate regulation remains a worthwhile research direction.

\section{Method}
\label{sec:method}

\subsection{Overview}
\label{sec:method_overview}

\begin{figure}[!t]
  \centering
  \includegraphics[width=\textwidth,keepaspectratio]{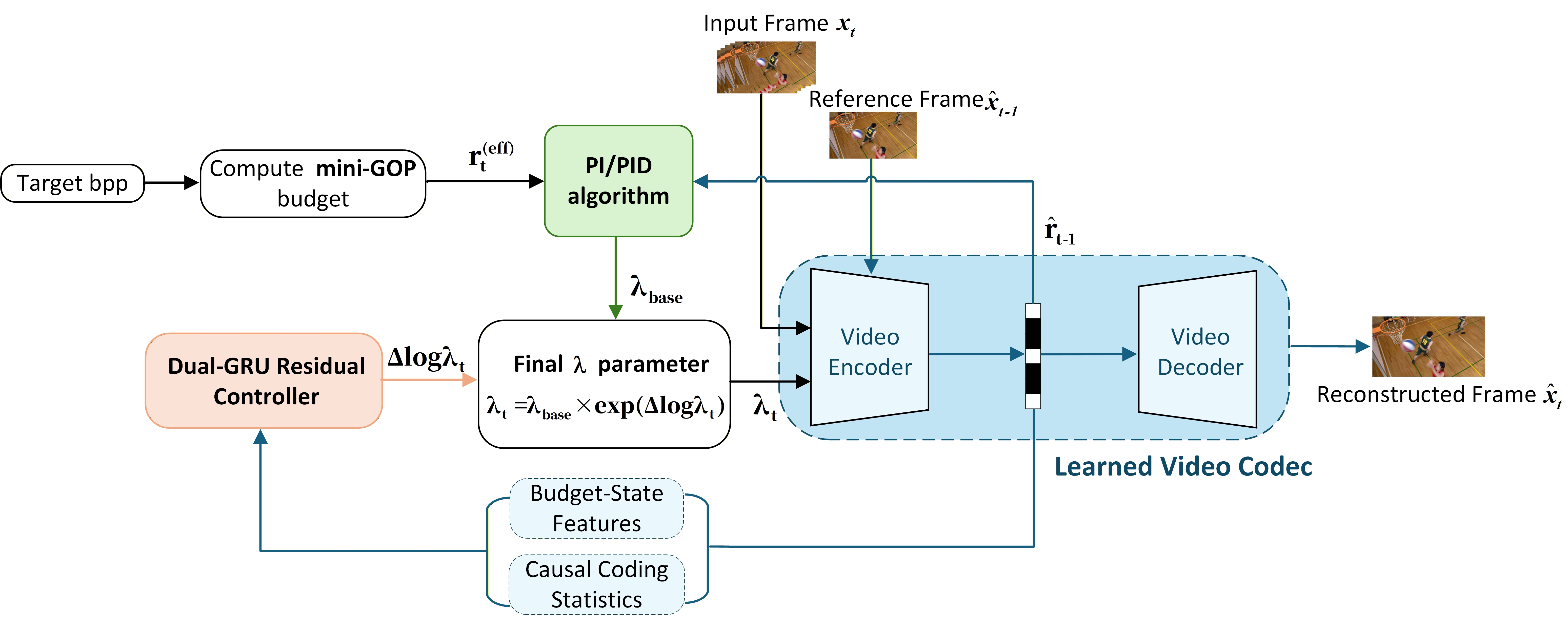}
 \caption{Overall framework of the proposed feedback-driven rate control method for learned video compression. A feedback controller first generates the base control signal $\lambda_t^{(\mathrm{base})}$ for target bitrate tracking, and the dual-GRU-based RD-optimized adjustment controller further predicts a bounded residual $\Delta_{\mathrm{GRU}} \log \lambda_t$ from budget-state features and causal coding statistics to refine the final coding parameter.}
  \label{fig:framework}
\end{figure}

This work focuses on the following problem in learned video compression: how to achieve stable online regulation under a given target bitrate constraint, while further improving the rate--distortion efficiency of inter-frame bit allocation under local budget constraints. Different from conventional fixed-$\lambda$ coding, target bitrate-driven online control not only requires the encoder to have a continuously adjustable bitrate response, but also requires the control mechanism to maintain stable target tracking under dynamically varying video content. Meanwhile, stable bitrate tracking alone is still insufficient to guarantee more effective inter-frame rate--distortion allocation. Therefore, bitrate controllability and allocation efficiency should be jointly modeled within a unified framework. The overall framework is illustrated in Fig.~\ref{fig:framework}.

Based on the above considerations, we divide the whole method into three tightly connected components. First, we introduce $\lambda$-conditioned modulation into a DCVC-based learned video coding framework, enabling a single model to cover multiple bitrate operating points and providing a continuously controllable rate--distortion interface for subsequent online regulation. Second, for target bitrate tracking, we formulate a unified PI/PID closed-loop control scheme in the log domain, which updates the base control variable $\lambda_t^{(\mathrm{base})}$ online according to the deviation between the target bitrate and the actual bitrate of the previous frame, thereby providing a stable baseline control for the encoding process. Furthermore, although the basic feedback controller can ensure the attainability and stability of the overall bitrate, it is still difficult to fully exploit the differences among frames in content complexity and reference importance under budget constraints. To address this issue, we further propose a budget-constrained RD-optimized adjustment controller, which predicts a frame-wise residual correction on top of the base control signal to improve inter-frame rate--distortion allocation within a local budget.

This adjustment controller jointly models two types of temporal information with different dynamic characteristics. One consists of Budget-State Features, which reflect budget consumption, remaining allocation, and accumulated deviation. The other consists of Causal Coding Statistics, which characterize current content variation and coding behavior. Since the former mainly captures slow-varying dynamics related to budget evolution, while the latter corresponds more closely to fast-varying dynamics caused by content complexity changes and coding statistic fluctuations, we employ two separate GRU branches to model them individually. Their representations are then adaptively integrated through a gating-based fusion mechanism, and the frame-wise adjustment term $\Delta_{\mathrm{GRU}} \log \lambda_t$ is finally predicted in the log domain. This adjustment term does not replace the base controller; instead, it refines the base control signal in the form of a bounded residual. In this way, the overall framework simultaneously achieves stable target bitrate tracking and local rate--distortion optimization under budget constraints. Accordingly, the final control variable for the current frame is jointly determined by the base control signal and the residual adjustment term, namely,
\begin{equation}
\lambda_t
=
\mathrm{clip}
\Big(
\lambda_t^{(\mathrm{base})}\exp(\Delta_{\mathrm{GRU}} \log \lambda_t),
\lambda_{\min},\lambda_{\max}
\Big).
\label{eq:lambda_compose}
\end{equation}
where $\lambda_t^{(\mathrm{base})}$ denotes the baseline control signal produced by the feedback controller, and $\Delta_{\mathrm{GRU}} \log \lambda_t$ is used to further refine frame-level allocation under budget constraints.

\subsection{Lambda-Modulated DCVC for Variable-Rate Coding}
\label{sec:method_lambda}

To enable a learned video encoder to cover multiple bitrate operating points within a single model, we first build a single-model multi-rate coding baseline on top of the DCVC framework based on $\lambda$-conditioned modulation. The core motivation is that an original fixed-$\lambda$ learned video compression model usually corresponds only to discrete rate--distortion operating points, and covering different target bitrates often requires training multiple separate models. Although this strategy can provide multiple RD points, it is difficult to offer a continuous, unified, and adjustable control interface for subsequent online rate control. In contrast, a single-model multi-rate mechanism can continuously change the coding behavior within the same encoder by adjusting the rate--distortion parameter $\lambda$, thereby establishing the necessary foundation for subsequent frame-level closed-loop control.

\begin{figure}[!t]
  \centering
  \includegraphics[width=\linewidth]{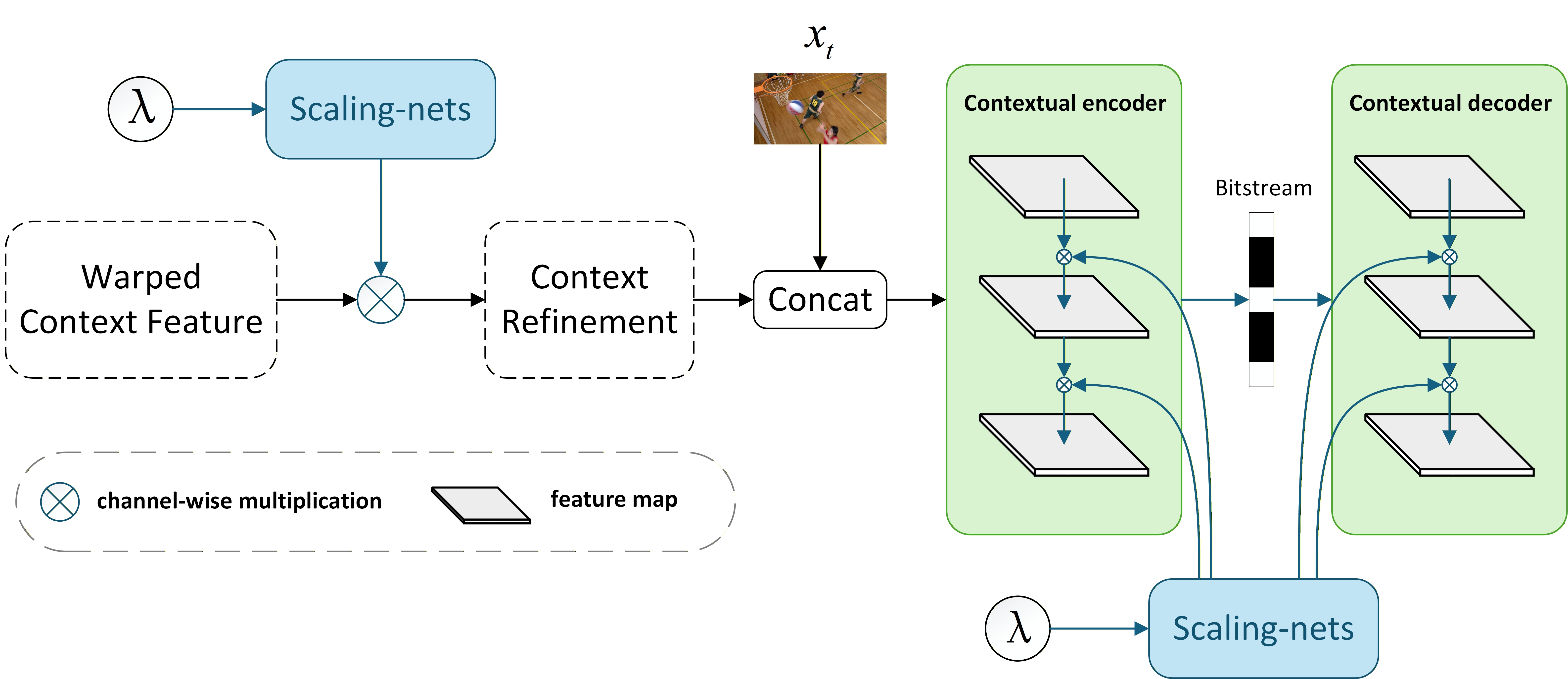}
  \caption{Illustration of the context-dependent path with $\lambda$-conditioned modulation introduced into the DCVC framework. We further extend the modulation mechanism to the spatiotemporal context modeling process: layer-wise feature modulation is inserted into the Contextual Encoder and Contextual Decoder, while an additional $\lambda$-conditioned scaling is applied to the warped context feature obtained by motion compensation, followed by Context Refinement to generate bitrate-dependent contextual representations.}
  \label{fig:my_label}
\end{figure}

Based on this idea, we adopt the modulation scheme proposed by Lin \emph{et al.}~\cite{lin2021deeply} as the starting point and adapt and extend it within the conditional coding architecture of DCVC. The core idea of the existing method is to introduce $\lambda$-conditioned modulation into multiple feature layers inside the auto-encoder, thereby enabling continuous control of the rate--distortion trade-off through channel-wise scaling. Different from applying modulation only to the compressed representation space, we further extend it to the spatiotemporal context modeling path of DCVC, so that the influence of $\lambda$ acts not only on latent representations but also enters the reference feature generation and conditional coding process. As shown in Fig.~\ref{fig:my_label}, we introduce layer-wise feature modulation into the Contextual Encoder and Contextual Decoder, while further applying $\lambda$-conditioned modulation to the warped context feature obtained by motion compensation, followed by Context Refinement to generate context representations associated with the target bitrate. Through this extension, the role of $\lambda$ is expanded from the conventional compressed representation space to the context modeling process, allowing it to participate in coding decisions at the spatiotemporal feature level.

Specifically, given a control parameter $\lambda$, we first normalize it and feed it into a set of lightweight modulation networks (Scaling-Nets) to generate channel-wise scaling factors for different layers. Let the intermediate feature representation at the $l$-th modulation position in the network be
$\mathbf{f}^{(l)} \in \mathbb{R}^{C_l \times H_l \times W_l}$,
where $C_l$, $H_l$, and $W_l$ denote the number of channels, height, and width of that layer, respectively. Let the corresponding modulation network be denoted by $\mathcal{M}^{(l)}(\cdot)$, which takes the scalar control parameter $\lambda$ as input and outputs a channel-wise scaling vector
$\mathbf{s}^{(l)} \in \mathbb{R}^{C_l}$. The modulation process can then be written as
\begin{equation}
\mathbf{s}^{(l)} = \mathcal{M}^{(l)}(\lambda),
\end{equation}
\begin{equation}
\tilde{\mathbf{f}}^{(l)} = \mathbf{s}^{(l)} \odot \mathbf{f}^{(l)},
\end{equation}
where $\odot$ denotes channel-wise multiplication, and $\mathbf{s}^{(l)}$ is broadcast along the spatial dimensions to $H_l \times W_l$, so that the feature response of each channel is scaled consistently.

In this work, such conditional modulation is applied at multiple key positions. First, layer-wise modulation is introduced at multiple feature stages inside the Contextual Encoder and Contextual Decoder, enabling $\lambda$ to directly affect the representation of contextual features during conditional coding. Second, $\lambda$-conditioned modulation is further applied to the warped context feature obtained by motion compensation, so that the reference feature becomes bitrate-aware before entering the subsequent Context Refinement module. Third, the modulated contextual feature, together with the current-frame information, is fed into the Contextual Encoder, allowing $\lambda$ to maintain a persistent influence on the spatiotemporal joint modeling process. In this way, $\lambda$ no longer acts only on the compressed representation itself, but instead runs through the entire process of context generation, conditional fusion, and compressed representation modeling.

During training, to enable a single model to cover multiple bitrate operating points, we jointly optimize the model over a predefined set of $\lambda$ values. Specifically, for each training sample, one control parameter is randomly sampled from a discrete $\lambda$ set and used as the conditional input for forward propagation. The model parameters are shared across different $\lambda$ conditions, so that the network learns to adapt to different rate--distortion trade-offs within a unified parameter space. The training objective follows the standard rate--distortion loss
\begin{equation}
\mathcal{L}_{\mathrm{RD}}
=
\lambda \cdot D(x,\hat{x})
+
R,
\end{equation}
where $D(\cdot)$ denotes the reconstruction distortion, $R$ denotes the bitrate estimated by the entropy model, and $\lambda$ controls the rate--distortion trade-off. Different from fixed-$\lambda$ training, $\lambda$ here is randomly sampled during training, allowing the same set of network parameters to cover multiple bitrate operating points. Considering that the reconstruction quality of I-frames has a significant effect on the subsequent chained coding of P-frames, we employ a strong I-frame model in both training and testing to reduce the interference of reference-frame errors on the stability of variable-rate modeling.

The resulting single-model multi-rate baseline obtained in this manner is denoted as DCVC-$\lambda$; the corresponding version constructed on DCVC-TCM in the same way is denoted as DCVC-TCM-$\lambda$. It should be emphasized that the significance of this step lies not only in obtaining a single-model baseline capable of covering multiple bitrate points, but also in establishing a continuous control interface directly coupled with rate--distortion behavior. All subsequent feedback control modules are built upon this unified interface: the PI/PID controller performs target bitrate tracking by updating $\lambda_t^{(\mathrm{base})}$ on a frame-by-frame basis, while the budget-constrained RD-optimized adjustment controller further predicts a bounded residual $\Delta_{\mathrm{GRU}} \log \lambda_t$ to optimize local bit allocation while maintaining control stability. In this way, $\lambda$-conditioned modulation extends the original fixed-rate codec into a single-model variable-rate codec that supports online feedback regulation, and provides a unified foundation for subsequent explicit feedback control and budget-constrained optimization.

\subsection{Log-domain PI/PID Closed-Loop Rate Control}
\label{sec:method_pid}

After establishing the variable-rate interface based on $\lambda$-conditioned modulation, the next step is to update $\lambda$ online on a frame-by-frame basis under a given target bitrate constraint, so that the output bitrate can stably track the target value. Different from traditional video coding, where the feedback loop is typically built around QP or buffer status, the control variable in learned video compression is the rate--distortion parameter $\lambda$, while the directly observable feedback signal is the bitrate $\hat r_t$ estimated by the entropy model. Therefore, the key issue is no longer to construct a mapping between discrete operating points, but rather to build a closed-loop regulation mechanism on top of a continuous control interface that can be dynamically corrected throughout the encoding process.

It should be emphasized that, although $\lambda$-conditioned modulation endows a single model with continuously adjustable bitrate response capability, it is still essentially an open-loop control mechanism: once $\lambda$ is given, the model simply produces the corresponding coding result under the current parameter condition, but it cannot guarantee that the actual sequence encoding process will stably reach the desired target bitrate. Due to the strong temporal dependency in learned video compression, motion intensity, texture complexity, and reference quality may vary continuously across adjacent frames. As a result, even under the same $\lambda$, the actual output bitrate of different frames may fluctuate significantly. Without an online feedback correction mechanism, such frame-wise deviations may accumulate over time and eventually cause the sequence-level bitrate to drift away from the target constraint. Based on this observation, we introduce an explicit closed-loop feedback control mechanism on top of the continuous control interface provided by $\lambda$-conditioned modulation, using the actual bitrate of the current frame as the feedback signal to update the control variable for the next frame online. Compared with open-loop approaches that rely on offline fitting or feed-forward prediction, feedback control can directly exploit the actual statistics observed during encoding, and is therefore better suited to maintaining stable target bitrate tracking in the presence of both content variation and model approximation errors.

Furthermore, we choose to perform feedback updates in the $\log \lambda$ domain rather than in the linear $\lambda$ domain. The reason is that the effect of $\lambda$ on bitrate is typically closer to a multiplicative relationship: at different operating points, the same absolute change in $\lambda$ may lead to drastically different bitrate variations, whereas the effect of relative changes is often more consistent. Therefore, updating the control variable in the log domain transforms multiplicative adjustment into an additive form, allowing the controller to maintain a more uniform response scale across different bitrate ranges and thereby improving control robustness across operating points.

Based on the above considerations, we define the bitrate error in the log domain as
\begin{equation}
e_t = \log \frac{\hat r_t}{r_t},
\label{eq:pid_err}
\end{equation}
where $r_t$ and $\hat r_t$ denote the target bitrate and actual bitrate of the current frame, respectively. This error definition characterizes the relative deviation of the current coding result from the target bitrate, and naturally matches the multiplicative adjustment mechanism in the log domain.

On this basis, we adopt a unified PI/PID form to update the base control signal in the $\log \lambda$ domain. The integral state is updated as
\begin{equation}
I_t = \mathrm{clip}(I_{t-1}+e_t,\,-I_{\max},\,I_{\max}),
\label{eq:pid_integral}
\end{equation}
the derivative term is defined as
\begin{equation}
d_t = e_t - e_{t-1},
\label{eq:pid_diff}
\end{equation}
and the control increment is written as
\begin{equation}
\Delta_{\mathrm{PI}} \log\lambda_t
=
-\left(k_p e_t + k_i I_t + k_d d_t\right),
\label{eq:pid_delta}
\end{equation}
which is further constrained by
\begin{equation}
\Delta_{\mathrm{PI}} \log\lambda_t
\leftarrow
\mathrm{clip}(\Delta_{\mathrm{PI}} \log\lambda_t,\,-\Delta_{\max},\,\Delta_{\max})
\label{eq:pid_delta_clip}
\end{equation}
to limit the per-frame update magnitude. Finally, the base control signal is updated in a multiplicative form as
\begin{equation}
\lambda_{t+1}^{(\mathrm{base})}
=
\mathrm{clip}
\left(
\lambda_t^{(\mathrm{base})}\exp(\Delta_{\mathrm{PI}} \log\lambda_t),
\lambda_{\min},\lambda_{\max}
\right).
\label{eq:pid_lambda_update}
\end{equation}

When the actual bitrate of the current frame is higher than the target bitrate, $e_t>0$, and the controller reduces $\log\lambda$, thereby decreasing the coding cost of subsequent frames. Conversely, when the current-frame bitrate is lower than the target value, the controller increases $\log\lambda$ so as to allocate more bits to subsequent frames. Here, the proportional term responds to the current deviation, the integral term compensates for long-term accumulated error, and the derivative term characterizes the trend of error variation.

Although we retain the full PI/PID formulation in the method description, in the frame-level control scenario of learned video compression, the observed bitrate often contains strong content-dependent fluctuations and model approximation errors. The derivative term is sensitive to such high-frequency noise, and may easily amplify inter-frame instability and introduce additional oscillation. Therefore, in our practical experiments, we adopt the more robust PI configuration as the base controller, i.e., we set $k_d=0$. Meanwhile, the clipping of the integral state and the control increment serve as anti-windup and aggressive-update suppression mechanisms, respectively, thereby improving the overall stability of the closed-loop controller.

From an implementation perspective, the algorithm operates in an online frame-by-frame manner: the current frame is first encoded under the given base control signal $\lambda_t^{(\mathrm{base})}$, yielding the actual bitrate estimate $\hat r_t$ from the entropy model; the controller then updates its internal states according to the current error and generates the base control signal $\lambda_{t+1}^{(\mathrm{base})}$ for the next frame. In our implementation, this closed-loop update is triggered only for P-frames, while I-frames are encoded independently and do not participate in the feedback loop. This design helps maintain the stability of reference-frame quality and avoids introducing unnecessary control perturbations at key frames.

\begin{algorithm}[t]
\caption{Online log-domain PI/PID update for frame-level rate control}
\label{alg:pi_pid}
\small
\begin{algorithmic}[1]
\Require Effective frame-level target bitrate sequence $\{r_t^{(\mathrm{eff})}\}_{t=1}^{T}$, initial control $\lambda_1^{(\mathrm{base})}$, controller gains $k_p,k_i,k_d$, bounds $\lambda_{\min},\lambda_{\max}$, $I_{\max}$, $\Delta_{\max}$
\State $I_0 \gets 0$, $e_0 \gets 0$
\For{$t=1$ to $T$}
    \If{$x_t$ is an I-frame}
        \State Encode $x_t$ independently with the I-frame codec
        \State $\lambda_{t+1}^{(\mathrm{base})} \gets \lambda_t^{(\mathrm{base})}$
        \State $I_t \gets I_{t-1}$, \quad $e_t \gets e_{t-1}$
        \State \textbf{continue}
    \EndIf
    \State Encode P-frame $x_t$ using $\lambda_t^{(\mathrm{base})}$
    \State Obtain entropy-estimated bitrate $\hat r_t$
    \State $e_t \gets \log(\hat r_t / r_t^{(\mathrm{eff})})$
    \State $I_t \gets \mathrm{clip}(I_{t-1}+e_t,\,-I_{\max},\,I_{\max})$
    \State $d_t \gets e_t - e_{t-1}$
    \State $\Delta_{\mathrm{PI}}\log\lambda_t \gets -(k_p e_t + k_i I_t + k_d d_t)$
    \State $\Delta_{\mathrm{PI}}\log\lambda_t \gets \mathrm{clip}(\Delta_{\mathrm{PI}}\log\lambda_t,\,-\Delta_{\max},\,\Delta_{\max})$
    \State $\lambda_{t+1}^{(\mathrm{base})} \gets
    \mathrm{clip}\!\left(
    \lambda_t^{(\mathrm{base})}\exp(\Delta_{\mathrm{PI}}\log\lambda_t),
    \lambda_{\min},\lambda_{\max}\right)$
\EndFor
\end{algorithmic}
\end{algorithm}

The operational flow of the online feedback controller is summarized in Algorithm~\ref{alg:pi_pid}, which clarifies the state-update order and the frame-wise triggering rule during inference. Under budget-constrained inference, the frame-level control target is given by $r_t^{(\mathrm{eff})}$.

In all experiments of this paper, the above general formulation reduces to a PI controller by setting $k_d=0$.

Overall, our algorithm progressively drives the encoding process toward the target bitrate at the sequence level by correcting accumulated bitrate deviations on a frame-by-frame basis, and provides a reliable starting point for subsequent budget-constrained rate--distortion allocation.

\subsection{Budget-Constrained RD-Optimized Adjustment Controller}

Although the log-domain PI algorithm can establish the attainability of the target bitrate and suppress temporally accumulated errors, its primary role is still to regulate stability according to bitrate deviation, rather than to explicitly optimize inter-frame bit allocation under budget constraints. It can determine whether the current bitrate deviates from the target and pull the encoding process back toward the target through feedback updates, but it cannot further answer the following question: under a given local budget constraint, how should bits be allocated across different frames to achieve better rate--distortion performance? In learned video compression, adjacent frames often differ significantly in motion intensity, texture complexity, and reference quality. Therefore, relying only on a uniform error-feedback mechanism usually leads to stable but not necessarily optimal local bit allocation.

\begin{figure}[!t]
  \centering
  \includegraphics[width=\textwidth]{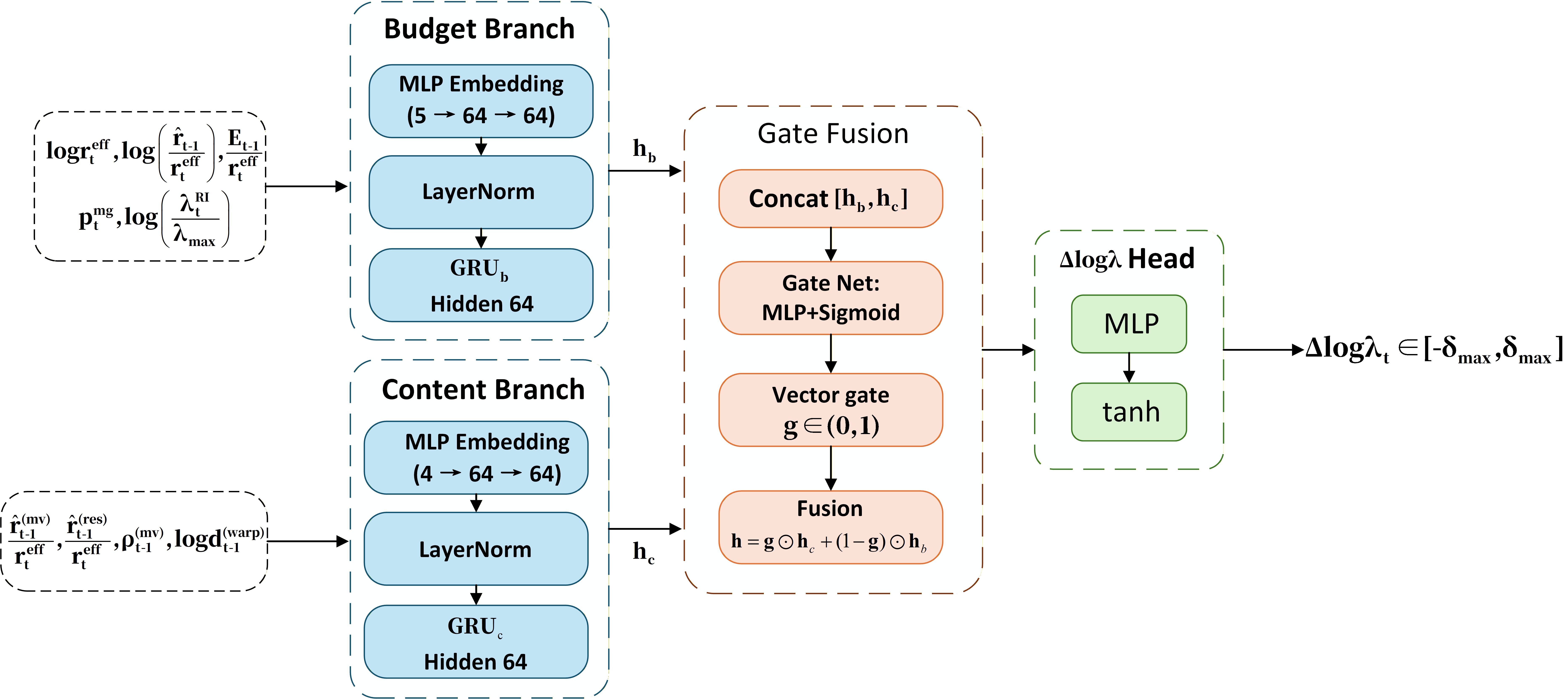}
  \caption{Budget-constrained RD-optimized adjustment controller. The controller takes budget-state features and causal coding statistics as inputs, predicts a bounded $\Delta_{\mathrm{GRU}} \log \lambda_t$, and refines the base $\lambda_t^{(\mathrm{base})}$ produced by the PI controller.}
  \label{fig:gru_controller}
\end{figure}

Based on this observation, we further introduce a budget-constrained RD-optimized adjustment controller on top of the base control signal $\lambda_t^{(\mathrm{base})}$ to predict a frame-wise correction term $\Delta_{\mathrm{GRU}} \log \lambda_t$, which is then composed with the base control signal to form the final control variable. In this way, the overall bitrate stability can be maintained while the rate--distortion efficiency under local budget constraints is further improved. This controller does not replace the base feedback controller; instead, it performs a bounded residual adjustment around it. Its role can therefore be understood as further optimizing frame-wise bit allocation within a local budget without compromising the stability of the closed-loop control process. The overall structure is shown in Fig.~\ref{fig:gru_controller}.

Under local budget constraints, frame-level bitrate allocation is essentially a sequential decision problem. The adjustment behavior for the current frame depends not only on the current target bitrate, but also on historical budget consumption, accumulated deviation, and previous coding states. Compared with feed-forward modeling based on single-frame inputs, recurrent structures are better suited to characterizing such cross-frame evolution. Considering the computational constraints of online encoding, we adopt the relatively lightweight GRU to model the temporal dynamics of the control state.

The controller takes two types of strictly causal input features: budget-state features $\mathbf{b}_t$ and coding-statistics features $\mathbf{c}_t$. The former describe the current system budget status and control context, such as the target bitrate, accumulated deviation, and the base control signal; the latter characterize the coding behavior and content complexity at the previous time step, such as the decomposition of motion and residual bitrates, motion sparsity, and motion compensation error. These two types of features differ not only semantically, but also in their dynamic properties: budget-state features mainly reflect cross-frame accumulation and budget evolution, and thus exhibit stronger slow-varying characteristics; coding-statistics features more directly reflect local content changes and coding-behavior fluctuations, and therefore exhibit more evident fast-varying characteristics. Modeling them separately is thus more beneficial for preserving their respective temporal representation capabilities.

Specifically, the budget-state features are defined as
\begin{equation}
\mathbf{b}_t =
\big[
\log r_t^{\mathrm{eff}},\;
\log(\hat r_{t-1}/r_t^{\mathrm{eff}}),\;
E_{t-1}/r_t^{\mathrm{eff}},\;
p_t,\;
\log(\lambda_t^{(\mathrm{base})}/\lambda_{\max})
\big],
\label{eq:budget_feat}
\end{equation}
and the coding-statistics features are defined as
\begin{equation}
\mathbf{c}_t =
\big[
\hat r^{(mv)}_{t-1}/r_t^{\mathrm{eff}},\;
\hat r^{(res)}_{t-1}/r_t^{\mathrm{eff}},\;
\rho^{(mv)}_{t-1},\;
\log d^{(\mathrm{warp})}_{t-1}
\big],
\label{eq:content_feat}
\end{equation}
where $r_t^{\mathrm{eff}}$ denotes the effective target bitrate assigned to frame $t$ after budget projection, which serves as the normalized reference rate for the current control step; $E_{t-1}$ denotes the accumulated signed budget deviation up to frame $t-1$, reflecting the difference between the consumed bits and the allocated budget in previous steps; and $p_t$ denotes the normalized progress of the current frame within the current mini-GOP. Moreover, $\hat r_{t-1}^{(mv)}$ and $\hat r_{t-1}^{(res)}$ denote the entropy-estimated motion and residual bitrates of frame $t-1$, respectively; $\rho_{t-1}^{(mv)}$ denotes a motion-related sparsity indicator computed from the motion representation; and $d_{t-1}^{(\mathrm{warp})}$ denotes the motion-compensation error of the previous frame measured on the warped prediction. Here, $\lambda_t^{(\mathrm{base})}$ is the base control signal generated by the underlying PI controller, and $\lambda_{\max}$ is the upper bound of the allowable control range. All these quantities are available at the current encoding step from the coding results of the current or previous frames only, thereby ensuring strict causality during inference.

To simultaneously model the temporal dynamics of budget evolution and content variation, we adopt a dual-branch GRU structure. The two groups of features are first mapped by MLPs into a unified low-dimensional embedding space and normalized, and are then fed into two independent GRU branches to obtain hidden states $\mathbf{h}_b$ and $\mathbf{h}_c$, respectively. These two hidden states are then adaptively fused through a gating mechanism. Specifically, $\mathbf{h}_b$ and $\mathbf{h}_c$ are concatenated and fed into a Gate Net, implemented as a two-layer MLP with the structure $128 \rightarrow 64 \rightarrow 64$, followed by a sigmoid activation to generate channel-wise weights $\mathbf{g}\in(0,1)^{64}$. The fused representation is then obtained as
\begin{equation}
\mathbf{h}
=
\mathbf{g}\odot\mathbf{h}_c
+
(1-\mathbf{g})\odot\mathbf{h}_b .
\end{equation}
This gated fusion is not a simple feature concatenation. Instead, it allows the controller to adaptively adjust its focus between the two state representations according to the current budget pressure and the degree of content variation: when the budget constraint becomes tighter, the budget-state features should play a stronger role; when the budget is relatively relaxed but content complexity changes significantly, the coding-statistics features become correspondingly more important.

On this basis, the fused feature is used to predict the frame-wise adjustment term
\begin{equation}
\Delta_{\mathrm{GRU}} \log \lambda_t
=
\mathcal{F}_{\mathrm{ctrl}}(\mathbf{b}_t,\mathbf{c}_t),
\label{eq:gru_delta_general}
\end{equation}
where the output is constrained by $\tanh$ and scaled to the interval $[-\delta_{\max},\delta_{\max}]$ to limit the magnitude of control perturbation. The final control signal is then given by
\begin{equation}
\lambda_t
=
\mathrm{clip}
\Big(
\lambda_t^{(\mathrm{base})}
\exp(\Delta_{\mathrm{GRU}} \log \lambda_t),
\lambda_{\min},\lambda_{\max}
\Big)
\label{eq:gru_lambda_compose}
\end{equation}

It should be emphasized that this controller does not directly predict the final control parameter, but only performs a residual correction to the base control signal. Such a design has two benefits. First, the final control result always takes the output of the base feedback controller as the reference, thereby preserving the interpretability and stability brought by the explicit feedback mechanism. Second, by imposing a bounded constraint on the residual, the learned module is prevented from producing overly aggressive control perturbations. Therefore, the functional boundary of this module is clear: it is not intended to replace the PI algorithm in target bitrate tracking, but rather to further optimize frame-wise allocation within a local budget once stable target tracking has already been established.

To learn the above residual correction strategy, we freeze the codec backbone during training and optimize only the parameters of the adjustment controller. The training objective is not to directly regress an ideal control variable, but to indirectly learn frame-wise correction behavior through rate--distortion optimization under budget constraints. Specifically, for a training mini-GOP, the frame-wise correction term output by the controller is first composed with the base control signal $\lambda_t^{(\mathrm{base})}$ to obtain the final control parameter $\lambda_t$, which is then used to drive the frozen codec to encode and reconstruct the corresponding frame. On this basis, we adopt the following objective:
\begin{equation}
\mathcal{L}
=
w_{\mathrm{dist}}\,\mathcal{L}_{\mathrm{dist}}
+
w_{\mathrm{budget}}\,\mathcal{L}_{\mathrm{budget}}
+
w_{\mathrm{smooth}}\,\mathcal{L}_{\mathrm{smooth}},
\label{eq:train_loss_all}
\end{equation}
where
\begin{equation}
\mathcal{L}_{\mathrm{dist}}
=
\sum_{t \in \mathcal{G}} D(x_t,\hat{x}_t),
\end{equation}
\begin{equation}
\mathcal{L}_{\mathrm{budget}}
=
\left(\bar r_{\mathrm{mg}} - r_{\mathrm{mg}}^{*}\right)^2,
\end{equation}
\begin{equation}
\mathcal{L}_{\mathrm{smooth}}
=
\sum_{t \in \mathcal{G}}
(\Delta_t - \Delta_{t-1})^2,
\end{equation}
where $\mathcal{G}$ denotes the set of frames in the current mini-GOP, $\bar r_{\mathrm{mg}}$ denotes the average bitrate of this mini-GOP, $r_{\mathrm{mg}}^{*}$ denotes the corresponding target budget, and $\Delta_t \equiv \Delta_{\mathrm{GRU}} \log \lambda_t$. These three loss terms respectively constrain reconstruction quality, local budget consistency, and control-trajectory smoothness, thereby enabling the controller to learn a more reasonable temporal bit-allocation strategy while satisfying the budget constraint.

\subsection{Training and Inference Strategy}
\label{sec:method_train_infer}

After training the single-model multi-rate codec, we further train the budget-constrained RD-optimized adjustment controller while freezing the codec backbone. The key motivation is that the base DCVC-$\lambda$ model already provides a continuously adjustable control interface, and the task of the subsequent control module is not to relearn the compression representation itself, but rather to learn how to exploit this interface for temporally adaptive regulation under budget constraints given a fixed codec response. Therefore, we adopt a training strategy that freezes the codec backbone and optimizes only the adjustment controller, while constructing a control process around mini-GOP budget constraints that is consistent with the inference stage.

During training, the DCVC codec backbone and its entropy model parameters are kept fixed, and only the adjustment controller is optimized. For the current frame, the base PI controller first produces $\lambda_t^{(\mathrm{base})}$, after which the adjustment controller predicts the frame-wise correction term $\Delta_{\mathrm{GRU}} \log \lambda_t$, and the final control signal $\lambda_t$ is composed according to Eq.~\eqref{eq:gru_lambda_compose}. It should be noted that the training objective is not to directly regress an ideal control parameter, but to learn how residual correction should be performed around the base control signal under local budget constraints. To avoid a mismatch between the training target and the actual reachable range of the current codec, we adopt a target-budget construction strategy based on pre-encoding: a $\lambda_{\mathrm{pre}}$ is randomly sampled from a predefined $\lambda$ set, and the frozen codec is used to perform one forward encoding pass on the designated frame to obtain the corresponding bitrate. This bitrate is then combined with the base PI control logic to construct the target mini-GOP budget $r_{\mathrm{mg}}^{*}$. Since this budget is directly derived from the true response of the frozen codec, it helps reduce the scale mismatch between training and practical inference, and enables the controller to learn an adjustment strategy that better matches the actual reachable operating range.

On this basis, the optimization objective of the controller consists of three parts: reconstruction quality, budget consistency, and control smoothness, whose definitions have been given in the previous subsection. During training, the controller takes local budget-state features and coding statistics at the mini-GOP level as input, and learns to output frame-wise residual correction terms that both satisfy the local budget constraint and improve inter-frame rate--distortion allocation. In this way, the stable baseline provided by the base PI controller and the local optimization capability learned by the adjustment controller are jointly integrated within a unified training framework.

During inference, we adopt the same local budget-driven mechanism as in training. Since the sequence-level target bitrate cannot be directly used for frame-wise regulation, it must first be converted into the available budget for the current mini-GOP. For the current mini-GOP, the budget is defined as
\begin{equation}
R_{\mathrm{mg}}
=
\left(
\frac{
R_s \cdot (N_{\mathrm{coded}} + SW) - \hat R
}{
SW
}
\right)\cdot N_m,
\label{eq:rmg_formula}
\end{equation}
where $R_{s}$ denotes the target average sequence bitrate, $N_{coded}$ denotes the number of frames that have already been
 encoded, $\bar{R}$ denotes the total accumulated consumed bits, $N_{m}$ denotes the number of frames in the current mini-GOP, and $SW$ denotes the smoothing-window length. Furthermore, let the remaining budget of the current mini-GOP be
\begin{equation}
R_{\mathrm{rem}} = R_{\mathrm{mg}} - R_{\mathrm{spent}},
\label{eq:rmg_rem}
\end{equation}
then the effective frame-level target bitrate is defined as
\begin{equation}
r_t^{(\mathrm{eff})}
=
\mathrm{clip}
\left(
\frac{R_{\mathrm{rem}}}{N_{\mathrm{rem}}},
\, r_{\min},\, r_{\max}
\right),
\label{eq:target_eff}
\end{equation}
where $N_{\mathrm{rem}}$ denotes the number of remaining frames within the current mini-GOP. This quantity can be regarded as the instantaneous target bitrate allocated to the current frame under the local budget constraint.

Based on the above budget update, the frame-wise control process during inference proceeds as follows: first, according to $r_t^{(\mathrm{eff})}$ and the current bitrate feedback, the base PI controller updates $\lambda_t^{(\mathrm{base})}$; next, the adjustment controller outputs a bounded correction term $\Delta_{\mathrm{GRU}} \log \lambda_t$; finally, the two are composed into the final control parameter $\lambda_t$, which is used to encode the current frame. Accordingly, the entire online control process can be summarized as follows: mini-GOP budget update $\rightarrow$ effective target bitrate computation $\rightarrow$ base control signal update $\rightarrow$ residual correction prediction $\rightarrow$ current-frame encoding.

Overall, this training and inference strategy ensures that the controller always operates under codec responses and budget constraints that are consistent with practical inference. The base PI controller is responsible for maintaining the attainability and stability of the target bitrate, while the adjustment controller further optimizes frame-wise bit allocation within the local budget on this basis, thereby integrating the above components into a complete feedback-driven variable-rate control framework.

\section{Experiments}
\label{sec:experiments}
\subsection{Experimental Settings}
\label{sec:exp_settings}

\paragraph{Datasets.}
We use the Vimeo-90k dataset~\cite{xue2019video} for training. For evaluation, experiments are conducted on UVG~\cite{mercat2020uvg} and the HEVC standard test sequences (Classes B, C, D, and E). These datasets cover content with different resolutions, motion characteristics, and texture complexities, making them suitable for evaluating the effectiveness of variable-rate control and inter-frame bit allocation. Unless otherwise specified, during testing we encode the first 96 frames of each sequence with a GOP size of $G=32$. I-frames are independently encoded as fixed reference frames, while the remaining P-frames are adjusted and encoded frame by frame under the proposed online rate control mechanism. For a given target bitrate point, the encoder generates the control variable online and encodes the entire sequence accordingly, from which the RD curves and bitrate tracking results are obtained.

\paragraph{Evaluation Metrics.}
The experiments are mainly evaluated from two aspects: rate--distortion performance and rate control accuracy. Rate--distortion performance is characterized by PSNR--bpp curves, and the overall compression performance is compared using the BD-rate metric~\cite{bjontegaard2001calculation}. Rate control accuracy is measured by the average relative bitrate error
\begin{equation}
\Delta R(\%) = \frac{|\hat r-r^{*}|}{r^{*}} \times 100\%.
\label{eq:delta_r}
\end{equation}
where $r^{*}$ denotes the target average P-frame bitrate at a given target point, and $\hat r$ denotes the actual average P-frame bitrate obtained from sequence encoding. When necessary, frame-wise or mini-GOP budget statistics are also analyzed to assess local budget consistency.

\paragraph{Implementation Details.}
We use DCVC~\cite{li2021deep} and DCVC-TCM~\cite{sheng2022temporal} as the baseline models. Based on the modulation scheme of Lin \emph{et al.}~\cite{lin2021deeply}, we further extend $\lambda$-conditioned modulation to the context-dependent coding path of DCVC, and retrain the model to obtain the single-model multi-rate baseline DCVC-$\lambda$; the corresponding version on DCVC-TCM is constructed in the same manner.

In the experiments, we compare the following settings on both the DCVC and DCVC-TCM backbones:
1) the original fixed-$\lambda$ codec as the baseline;
2) the single-model multi-rate version with $\lambda$-conditioned modulation;
3) the log-domain PI algorithm built on top of it;
4) the complete model obtained by further introducing the budget-constrained RD-optimized adjustment controller on top of the PI algorithm.
Since this work focuses on the online rate regulation behavior in the inter-frame prediction chain, only P-frame bitrate is counted during evaluation. As all methods share the same I-frame encoder, this evaluation protocol more directly isolates and assesses the effect of the proposed control modules on the inter coding process. Unless otherwise specified, all rate-control-related comparisons in this paper are conducted under the same P-frame-only evaluation protocol.

For the budget-constrained RD-optimized adjustment controller, we adopt controller-only training, i.e., the codec backbone is frozen and only the controller parameters are optimized. This setting ensures that the controller learns a budget-aware adjustment strategy under a fixed codec response, rather than relearning the compression representation itself. During training, the Adam optimizer is used with a learning rate of $1\times10^{-4}$ and a batch size of 4. The controller is trained for 20 epochs in total, and a StepLR learning rate decay strategy is adopted, where the learning rate is multiplied by 0.5 every 5 epochs.

Although a unified PI/PID general form is adopted in the method description to present the feedback controller, all experiments in this paper use the log-domain PI configuration to obtain more robust bitrate regulation behavior. Specifically, the controller gains are set to $k_p=0.9$ and $k_i=0.05$, with $k_d=0$. The initial control parameter is set to $\lambda_0=1024$. To ensure control stability, $\lambda$ is constrained within the range $[32,4096]$, and the per-frame update magnitude satisfies $|\Delta_{\mathrm{PI}} \log \lambda_t| \le 0.30$. Meanwhile, the integral term is clipped to avoid oscillation caused by integral accumulation, with the clipping threshold set to $10.0$. These parameters are kept unchanged throughout all experiments. For the local budget calculation during inference, the mini-GOP size is set to $N_m=4$, and the smoothing window length is set to $SW=40$. The PI algorithm first produces the base control signal, after which the adjustment controller predicts $\Delta_{\mathrm{GRU}} \log \lambda_t$ and composes it with the base control signal to obtain the final coding parameter.

\begin{figure}[!t]
    \centering
    \includegraphics[width=\linewidth]{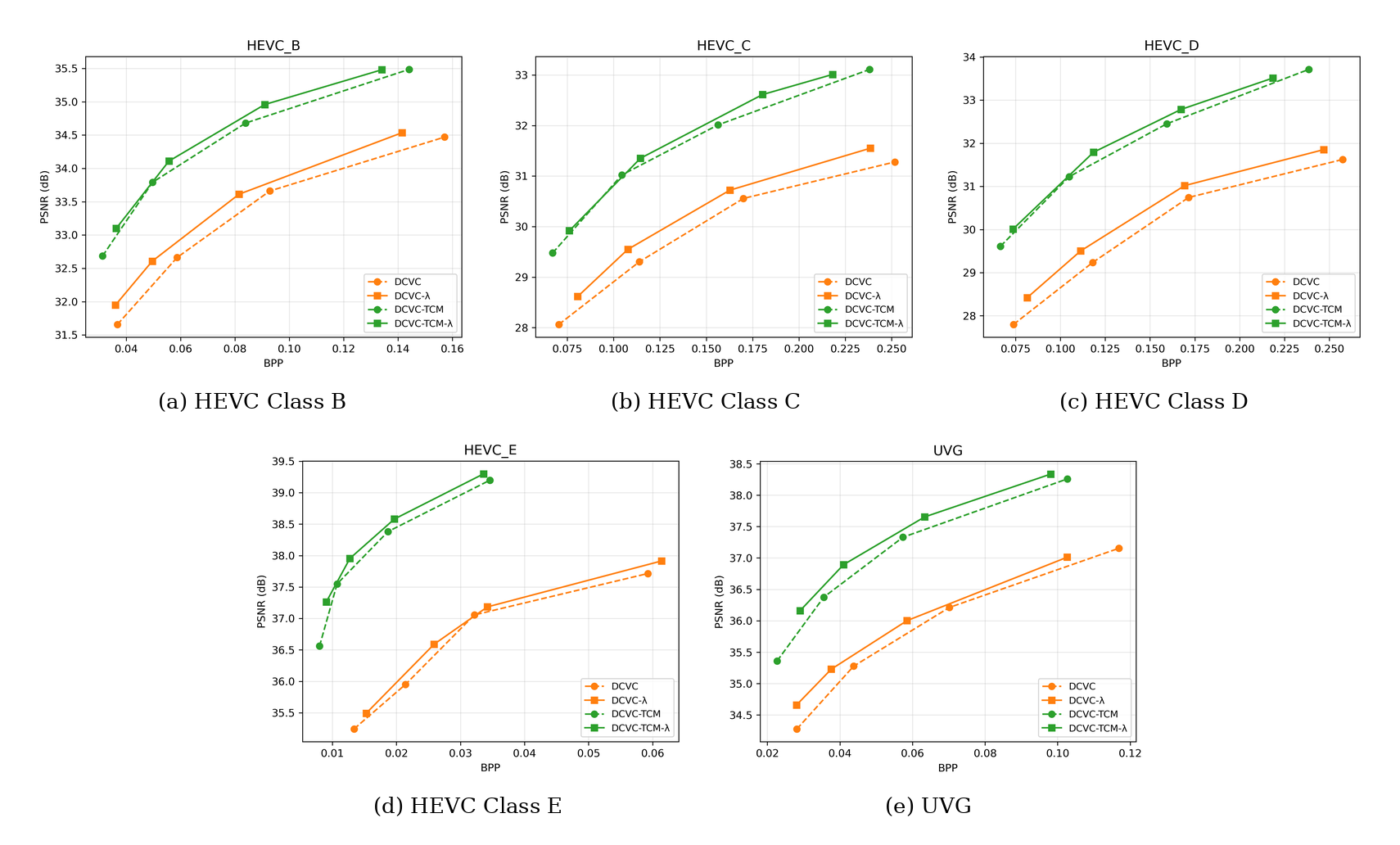}
    \caption{Rate-distortion comparison of the lambda-modulated single-model baseline on the HEVC Class B/C/D/E and UVG datasets.}
    \label{fig:rd_lambda}
\end{figure}

\subsection{Main Results}
\label{sec:exp_main_results}

This section evaluates the proposed method from three aspects. First, we verify the continuously adjustable capability of the single-model multi-rate interface. Second, we report the overall quantitative results of the proposed target-driven control framework, including both the PI baseline and the budget-constrained RD-optimized adjustment controller. Finally, we analyze the local budget alignment behavior and subjective reconstruction quality of the proposed method.

\subsubsection{Lambda-Modulated Variable-Rate Baseline}

We construct single-model multi-rate baselines on both the DCVC and DCVC-TCM frameworks. Fig.~\ref{fig:rd_lambda} shows the rate--distortion curves of the constructed single-model multi-rate baselines on the HEVC Class B/C/D/E and UVG test sets. It can be observed that the single-model scheme based on $\lambda$-conditioned modulation is able to cover a wide target bitrate range within one model, demonstrating good continuous bitrate adjustment capability. Furthermore, on the test sets shown, after extending the modulation mechanism to the context-dependent path and performing joint training, DCVC-$\lambda$ achieves better rate--distortion performance than the unmodulated baseline. This indicates that the $\lambda$ conditional information not only provides a continuous control interface, but also improves the compression representation capability at different bitrate points. However, since there still exists a mismatch between the actual output bitrate and the target bitrate under the same $\lambda$, the RD points of different methods do not strictly correspond to identical target bitrate conditions. Therefore, this baseline is mainly used to demonstrate the continuous bitrate adjustment capability of a single model, rather than serving as the primary BD-Rate comparison target in the subsequent target-bitrate-driven experiments.

\subsubsection{Overall Quantitative Results of Target-Driven Rate Control}

\begin{figure}[!t]
    \centering
    \includegraphics[width=\textwidth]{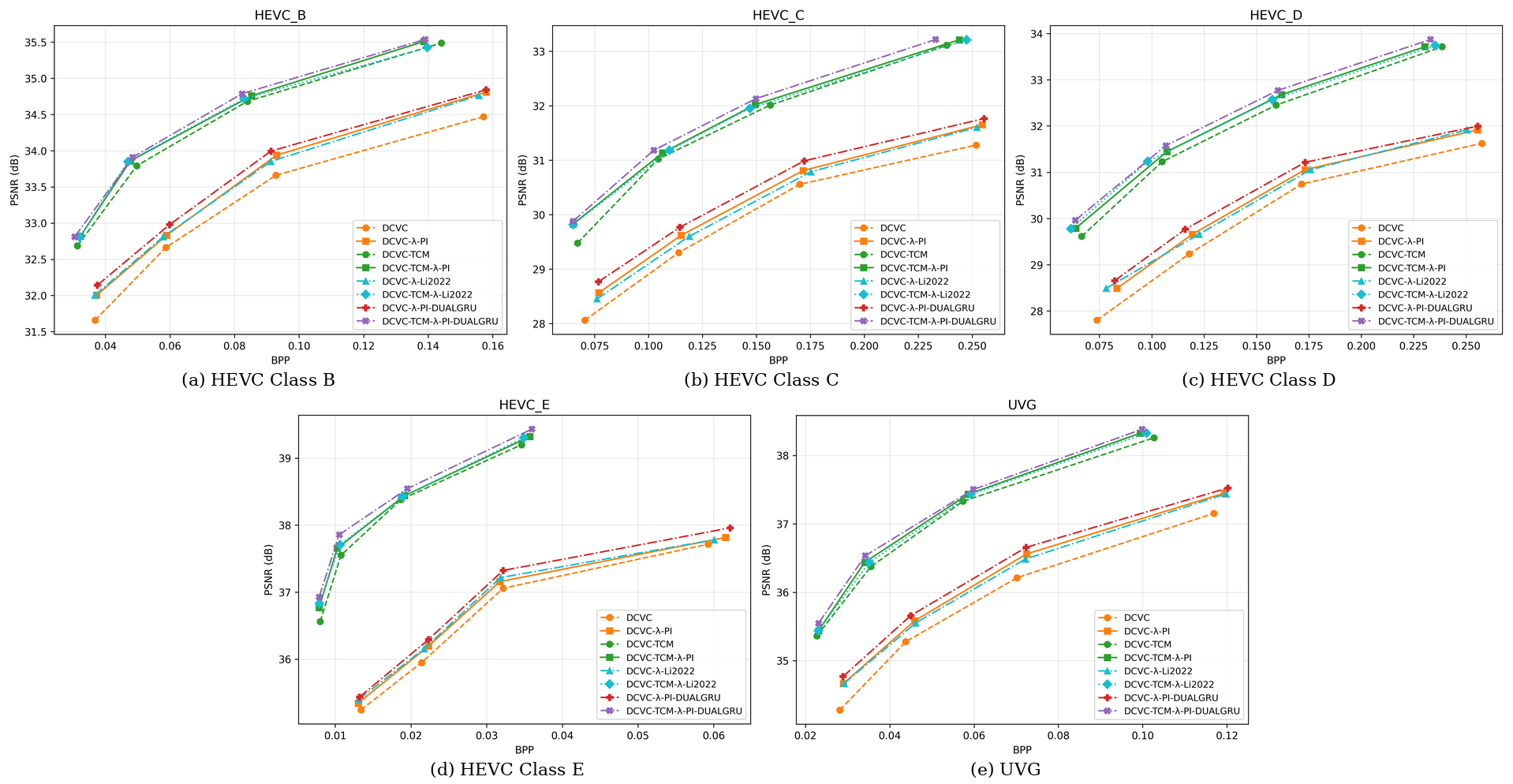}
    \caption{Overall rate--distortion comparison on the HEVC Class B/C/D/E and UVG datasets. The compared methods include the fixed-$\lambda$ baselines, Li2022, the proposed PI closed-loop control, and the proposed PI + DualGRU on both the DCVC and DCVC-TCM backbones.}
    \label{fig:pid_alignment}
\end{figure}

After establishing the continuously adjustable single-model multi-rate interface, we next evaluate the overall quantitative performance of the proposed target-driven control framework. Under the P-frame-only evaluation protocol described in Sec.~\ref{sec:exp_settings}, we first examine the target bitrate tracking capability of the PI baseline, and then evaluate the additional gains brought by the budget-constrained RD-optimized adjustment controller. The corresponding overall RD comparison is finally summarized in Fig.~\ref{fig:pid_alignment}.

\begin{figure}[ht]
    \centering
    \includegraphics[width=0.8\linewidth]{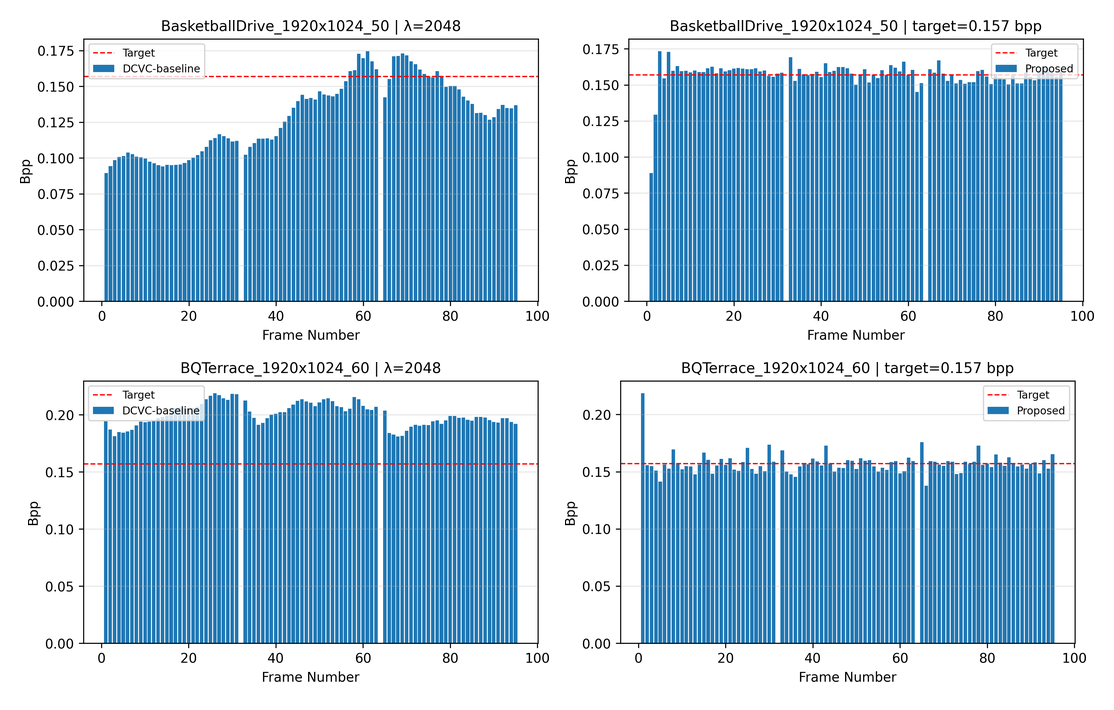}
    \caption{Frame-wise bitrate comparison on two representative sequences under a common target bitrate. For each sequence, the left panel shows the DCVC-$\lambda$ baseline and the right panel shows the proposed PI control result. The proposed method tracks the target bitrate more stably through online $\lambda$ updates.}
    \label{fig:pid_case_two_seq}
\end{figure}

Fig.~\ref{fig:pid_case_two_seq} illustrates the frame-wise rate control behavior of the PI controller on representative video sequences. Compared with the fixed-$\lambda$ baseline, the proposed method is able to keep the output bitrate distributed more stably around the target value by updating $\lambda$ online, thereby significantly improving target bitrate tracking performance.

\begin{table}[!t]
\centering
\caption{Main results of PI rate control. Negative BD-Rate indicates bitrate saving over the corresponding baseline, and $\Delta R$ denotes the average relative bitrate error (\%).}
\label{tab:main_pid_results}

\renewcommand{\arraystretch}{1.35}
\setlength{\tabcolsep}{5pt}

\resizebox{0.98\linewidth}{!}{%
\begin{tabular}{lcccccccccccc}
\toprule
& \multicolumn{2}{c}{HEVC B} 
& \multicolumn{2}{c}{HEVC C}
& \multicolumn{2}{c}{HEVC D}
& \multicolumn{2}{c}{HEVC E}
& \multicolumn{2}{c}{UVG}
& \multicolumn{2}{c}{Average} \\
\cmidrule(lr){2-3}\cmidrule(lr){4-5}\cmidrule(lr){6-7}\cmidrule(lr){8-9}\cmidrule(lr){10-11}\cmidrule(lr){12-13}
Method & BD-Rate(\%) & $\Delta R(\%)$ & BD-Rate(\%) & $\Delta R(\%)$ & BD-Rate(\%) & $\Delta R(\%)$ & BD-Rate(\%) & $\Delta R(\%)$ & BD-Rate(\%) & $\Delta R(\%)$ & BD-Rate(\%) & $\Delta R(\%)$ \\
\midrule
DCVC-$\lambda$-Li2022 vs DCVC
& -10.92 & 0.99
& -9.26  & 6.01
& \textbf{-10.70} & \textbf{3.78}
& \textbf{-8.76}  & \textbf{1.33}
& -10.24 & 3.77
& \textbf{-9.98}
& 3.18 \\

DCVC-$\lambda$-PI vs DCVC
& \textbf{-11.24} & \textbf{0.73}
& \textbf{-9.65}  & \textbf{3.13}
& -9.66  & 3.98
& -6.65  & 3.06
& \textbf{-11.91} & \textbf{3.49}
& -9.82
& \textbf{2.88} \\

DCVC-TCM-$\lambda$-Li2022 vs DCVC-TCM
& -5.06 & 3.07
& -3.09  & 4.49
& \textbf{-6.27} & 4.20
& \textbf{-5.69}  & \textbf{1.11}
& -3.34  & \textbf{1.93}
& -4.69
& \textbf{2.96} \\

DCVC-TCM-$\lambda$-PI vs DCVC-TCM
& \textbf{-5.18} & \textbf{3.04}
& \textbf{-4.04}  & \textbf{2.98}
& -5.44  & \textbf{2.89}
& -5.67  & 3.03
& \textbf{-5.15}  & 2.80
& \textbf{-5.10}
& 2.95 \\
\bottomrule
\end{tabular}%
}
\end{table}

Table~\ref{tab:main_pid_results} reports the main quantitative results of the PI controller. For the DCVC backbone, DCVC-$\lambda$-PI achieves an average BD-Rate improvement of $-9.82\%$ over HEVC B/C/D/E and UVG, with an average relative bitrate error of $2.88\%$. For the stronger DCVC-TCM backbone, DCVC-TCM-$\lambda$-PI achieves an average BD-Rate improvement of $-5.10\%$, with an average relative bitrate error of $2.95\%$. These results show that, after transferring explicit PI feedback control into the $\lambda$ domain of learned video compression, the encoder is able to stably approach the target bitrate under the single-model multi-rate setting.

Table~\ref{tab:main_pid_results} also provides a comparison with Li2022, which adapts the traditional R-D-$\lambda$ modeling-based rate control paradigm to learned end-to-end video compression. On the DCVC backbone, the proposed method achieves a rate control accuracy comparable to that of Li2022 in terms of average $\Delta R$ (with $2.88\%$ and $3.18\%$, respectively), while obtaining better BD-Rate performance on HEVC B, HEVC C, and UVG, and showing an overall average BD-Rate that is close to that of Li2022. On the stronger DCVC-TCM backbone, the proposed method likewise achieves a similar average $\Delta R$ to Li2022, while obtaining better average BD-Rate performance ($-5.10\%$ vs. $-4.69\%$). These results indicate that explicit PI feedback regulation can achieve rate control accuracy comparable to existing R-D-$\lambda$ modeling methods in learned video compression, while delivering better rate--distortion performance in some settings.

\begin{table}[!t]
\centering
\caption{Comparison between the PI baseline and the proposed adjustment controller. BD-Rate is reported relative to the PI baseline, and $\Delta R$ denotes the absolute average relative bitrate error (\%).}
\label{tab:main_gru_results}

\renewcommand{\arraystretch}{1.35}
\setlength{\tabcolsep}{5pt}

\resizebox{0.98\linewidth}{!}{%
\begin{tabular}{lcccccccccccc}
\toprule
& \multicolumn{2}{c}{HEVC B} 
& \multicolumn{2}{c}{HEVC C}
& \multicolumn{2}{c}{HEVC D}
& \multicolumn{2}{c}{HEVC E}
& \multicolumn{2}{c}{UVG}
& \multicolumn{2}{c}{Average} \\
\cmidrule(lr){2-3}\cmidrule(lr){4-5}\cmidrule(lr){6-7}\cmidrule(lr){8-9}\cmidrule(lr){10-11}\cmidrule(lr){12-13}
Method & BD-Rate(\%) & $\Delta R(\%)$ & BD-Rate(\%) & $\Delta R(\%)$ & BD-Rate(\%) & $\Delta R(\%)$ & BD-Rate(\%) & $\Delta R(\%)$ & BD-Rate(\%) & $\Delta R(\%)$ & BD-Rate(\%) & $\Delta R(\%)$ \\
\midrule
DCVC-$\lambda$-PI
& 0 & 0.73
& 0 & 3.13
& 0 & 3.98
& 0 & 3.06
& 0 & 3.49
& 0 & 2.88 \\

DCVC-$\lambda$-PI + DualGRU Adjustment Controller
& \textbf{-4.77} & \textbf{0.20}
& \textbf{-6.16} & \textbf{2.66}
& \textbf{-5.66} & \textbf{3.23}
& \textbf{-6.75} & \textbf{2.45}
& \textbf{-5.11} & \textbf{2.09}
& \textbf{-5.69} & \textbf{2.13} \\

\midrule

DCVC-TCM-$\lambda$-PI
& 0 & 3.04
& 0 & 2.98
& 0 & 2.89
& 0 & 3.03
& 0 & 2.80
& 0 & 2.95 \\

DCVC-TCM-$\lambda$-PI + DualGRU Adjustment Controller
& \textbf{-2.98} & \textbf{2.34}
& \textbf{-4.53} & \textbf{2.68}
& \textbf{-4.59} & \textbf{2.01}
& \textbf{-7.15} & \textbf{1.85}
& \textbf{-3.18} & \textbf{2.31}
& \textbf{-4.49} & \textbf{2.24} \\
\bottomrule
\end{tabular}%
}
\end{table}

Table~\ref{tab:main_gru_results} further reports the gains obtained by introducing the budget-constrained RD-optimized adjustment controller on top of the PI baseline. Compared with the corresponding PI-based controller, the complete model achieves an average BD-Rate improvement of $-5.69\%$ on HEVC B/C/D/E and UVG with the DCVC-$\lambda$ backbone, and a further average improvement of $-4.49\%$ with the DCVC-TCM-$\lambda$ backbone. Meanwhile, the average $\Delta R$ is also reduced from $2.88\%$ to $2.13\%$, and from $2.95\%$ to $2.24\%$, respectively. The further reduction in $\Delta R$ mainly comes from the local frame-wise correction of $\lambda$ performed by the adjustment controller under the mini-GOP budget constraint, which helps reduce accumulated budget error. These results indicate that the adjustment controller achieves more effective local bit-allocation optimization while preserving the budget constraint, thereby bringing additional rate--distortion gains.

Fig.~\ref{fig:pid_alignment} further provides an overall RD comparison among the fixed-$\lambda$ baselines, Li2022, the proposed PI controller, and the proposed PI + DualGRU variant on the DCVC and DCVC-TCM backbones. It can be observed that the PI controller already achieves competitive RD performance under small bitrate error, while the DualGRU-based refinement further improves the RD curves under budget-constrained allocation. Taken together, Table~\ref{tab:main_pid_results}, Table~\ref{tab:main_gru_results}, and Fig.~\ref{fig:pid_alignment} show that the PI controller establishes a stable and competitive baseline for target bitrate tracking, while the budget-constrained RD-optimized adjustment controller further improves the overall rate--distortion performance on top of this baseline.

\subsubsection{Analysis of Budget Alignment and Subjective Quality}

To further understand the effect of the budget-constrained RD-optimized adjustment controller beyond the overall quantitative gains reported in Sec.~4.2.2, we analyze its local budget behavior and subjective reconstruction quality.

\begin{figure}[!ht]
    \centering
    \includegraphics[width=0.8\textwidth]{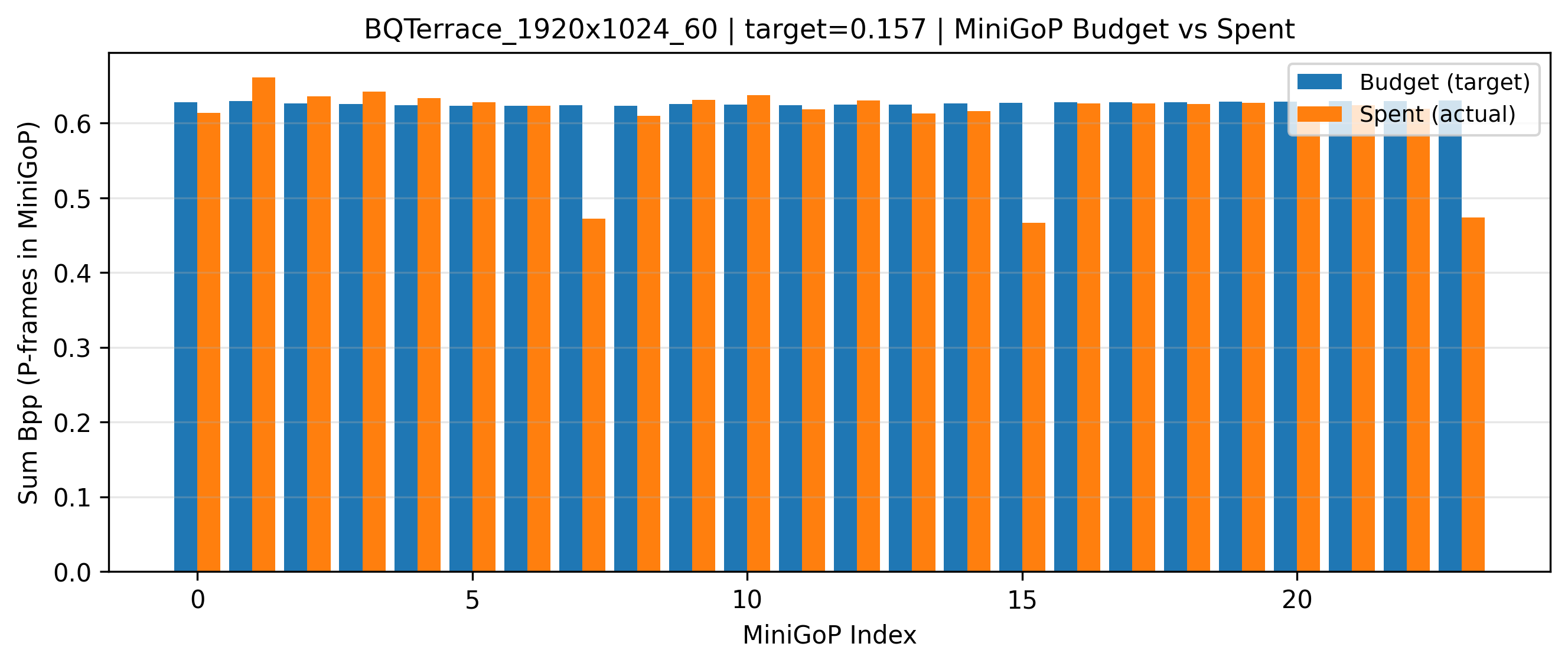}
    \caption{Example of mini-GOP budget alignment under the proposed budget-constrained adjustment controller.}
    \label{fig:minigop_alignment}
\end{figure}

\begin{figure}[!t]
    \centering

    \newcommand{\imgw}{0.205\textwidth}
    \newcommand{\imgh}{2.75cm}

    \begin{minipage}[t]{\imgw}
        \centering
        \includegraphics[width=\linewidth,height=\imgh]{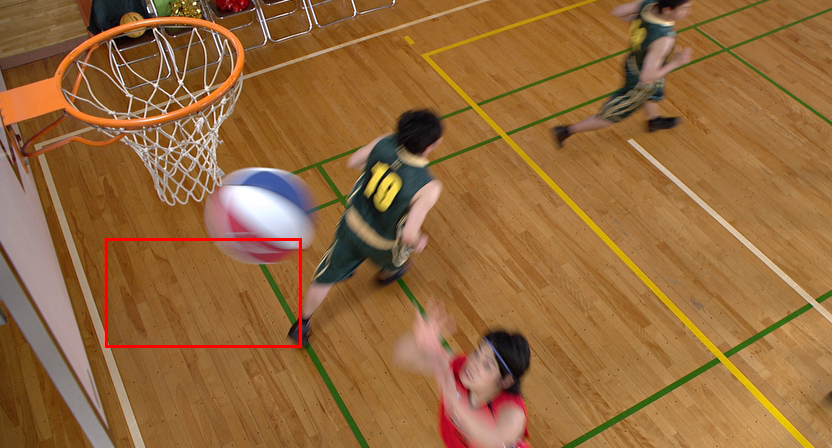}
        \vspace{-0.28em}
        \parbox[t][1.15em][t]{\linewidth}{\centering\scriptsize Original}
    \end{minipage}
    \begin{minipage}[t]{\imgw}
        \centering
        \includegraphics[width=\linewidth,height=\imgh]{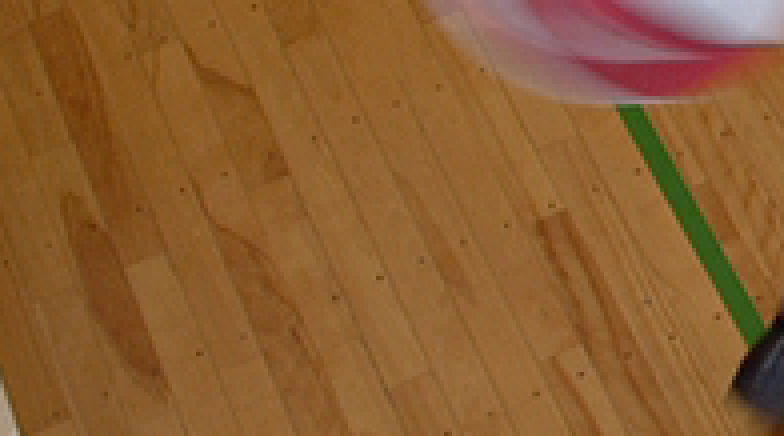}
        \vspace{-0.28em}
        \parbox[t][1.15em][t]{\linewidth}{\centering\scriptsize Ground Truth}
    \end{minipage}
    \begin{minipage}[t]{\imgw}
        \centering
        \includegraphics[width=\linewidth,height=\imgh]{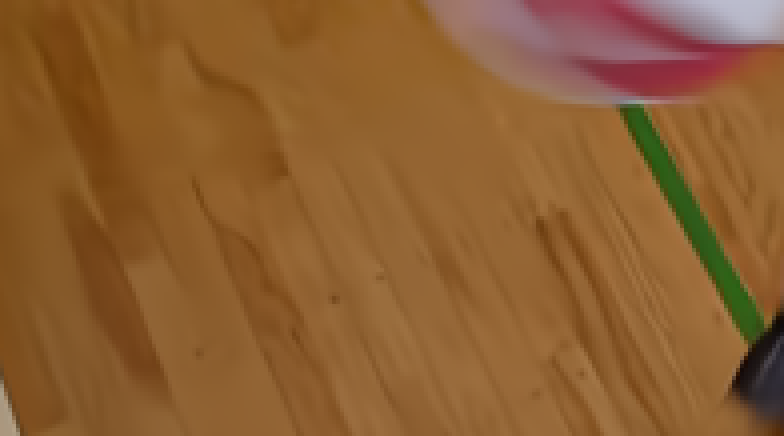}
        \vspace{-0.28em}
        \parbox[t][2.05em][t]{\linewidth}{\centering
            \scriptsize DCVC\\[-0.18em]
            \scriptsize (bpp: 0.1391, psnr: 33.90)}
    \end{minipage}
    \begin{minipage}[t]{\imgw}
        \centering
        \includegraphics[width=\linewidth,height=\imgh]{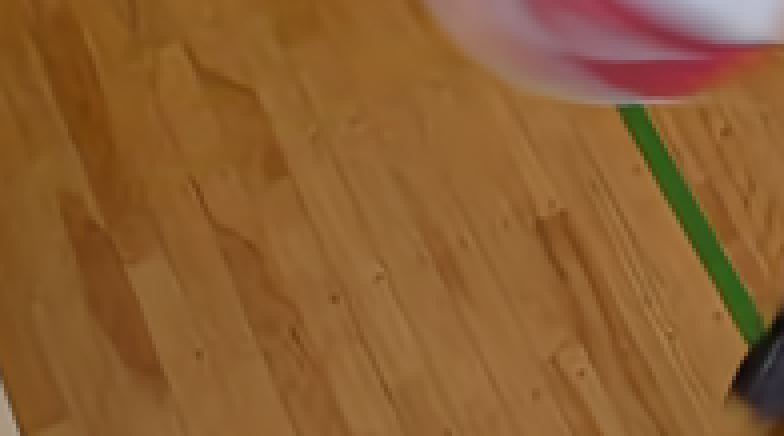}
        \vspace{-0.28em}
        \parbox[t][2.05em][t]{\linewidth}{\centering
            \scriptsize Ours\\[-0.18em]
            \scriptsize (bpp: 0.1158, psnr: 34.62)}
    \end{minipage}

    \vspace{-0.18em}
    {\small (a) HEVC Class C \textit{BasketballDrill}}

    \vspace{0.18em}

    \begin{minipage}[t]{\imgw}
        \centering
        \includegraphics[width=\linewidth,height=\imgh,
        trim=20.0cm 10cm 0cm 0.2cm,
            clip]{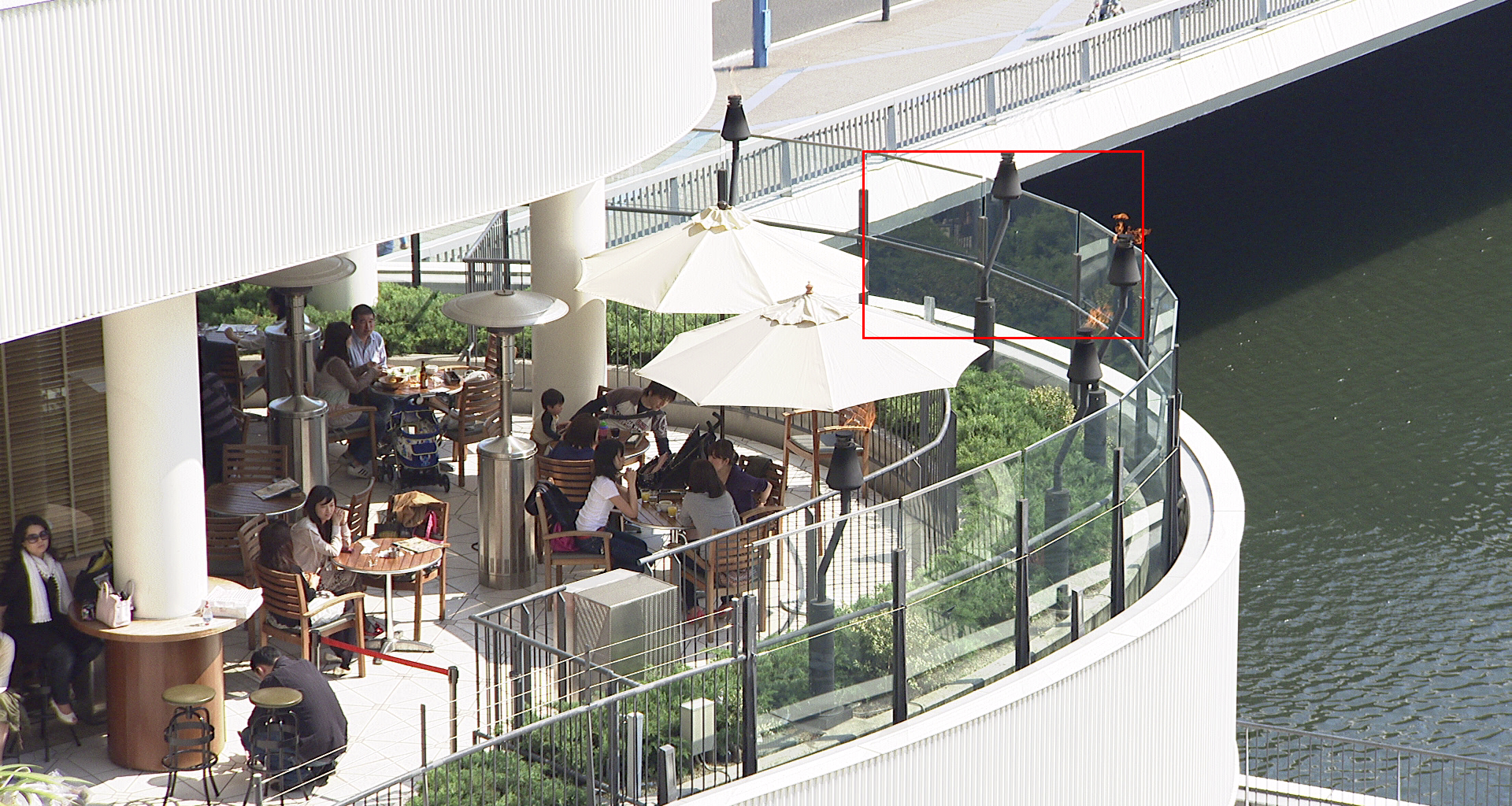}
        \vspace{-0.28em}
        \parbox[t][1.15em][t]{\linewidth}{\centering\scriptsize Original}
    \end{minipage}
    \begin{minipage}[t]{\imgw}
        \centering
        \includegraphics[width=\linewidth,height=\imgh]{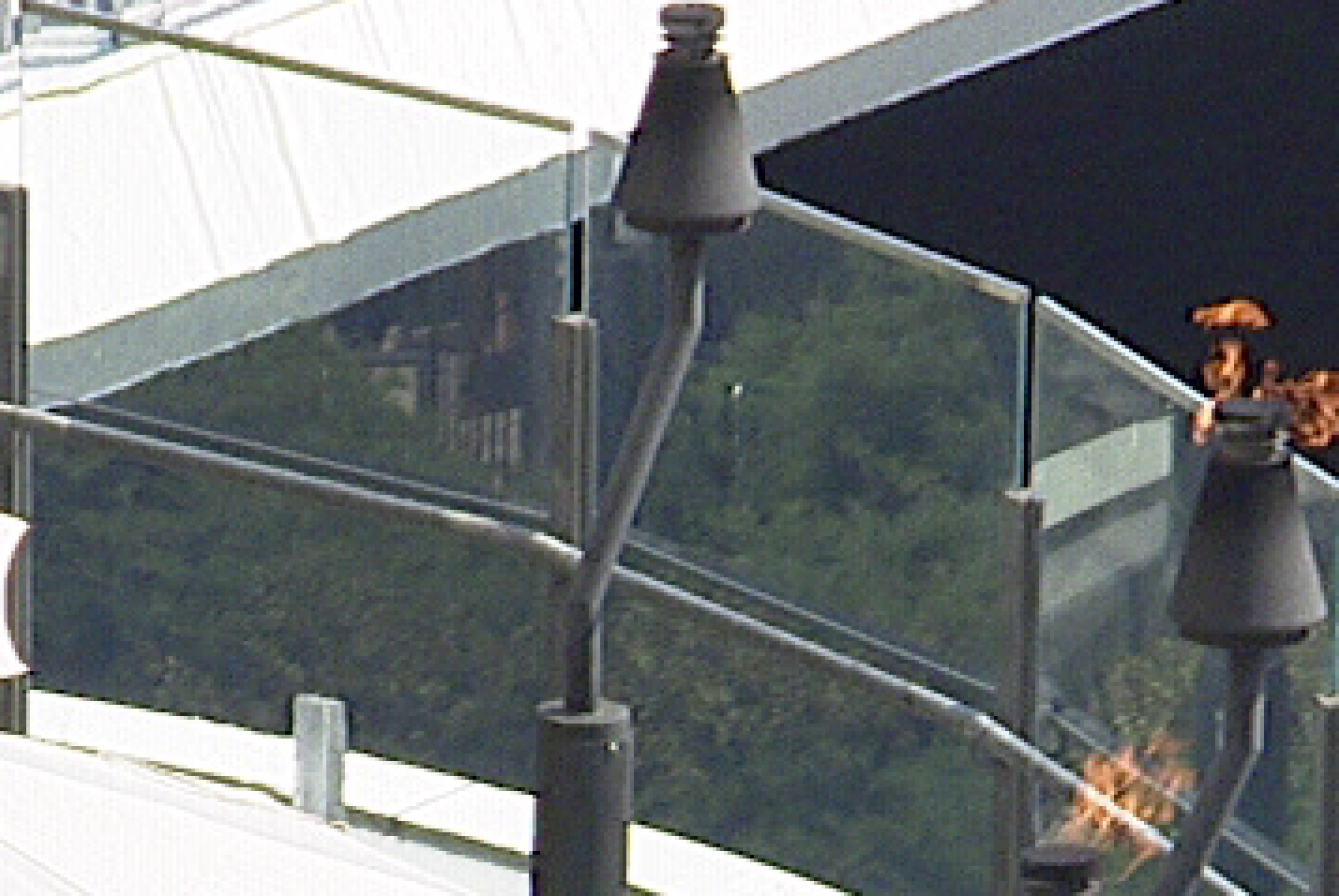}
        \vspace{-0.28em}
        \parbox[t][1.15em][t]{\linewidth}{\centering\scriptsize Ground Truth}
    \end{minipage}
    \begin{minipage}[t]{\imgw}
        \centering
        \includegraphics[width=\linewidth,height=\imgh]{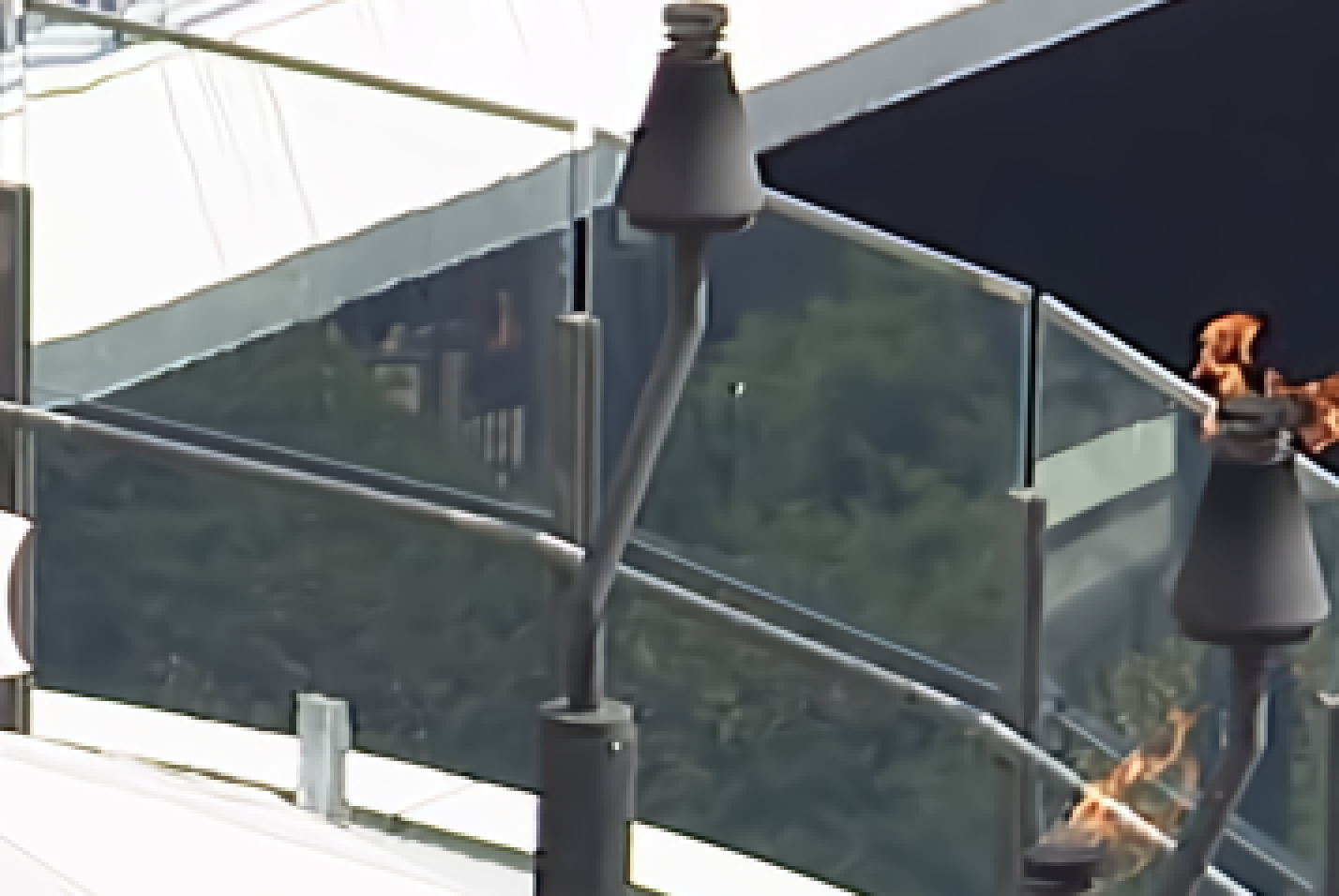}
        \vspace{-0.28em}
        \parbox[t][2.05em][t]{\linewidth}{\centering
            \scriptsize DCVC\\[-0.18em]
            \scriptsize (bpp: 0.1868, psnr: 32.78)}
    \end{minipage}
    \begin{minipage}[t]{\imgw}
        \centering
        \includegraphics[width=\linewidth,height=\imgh]{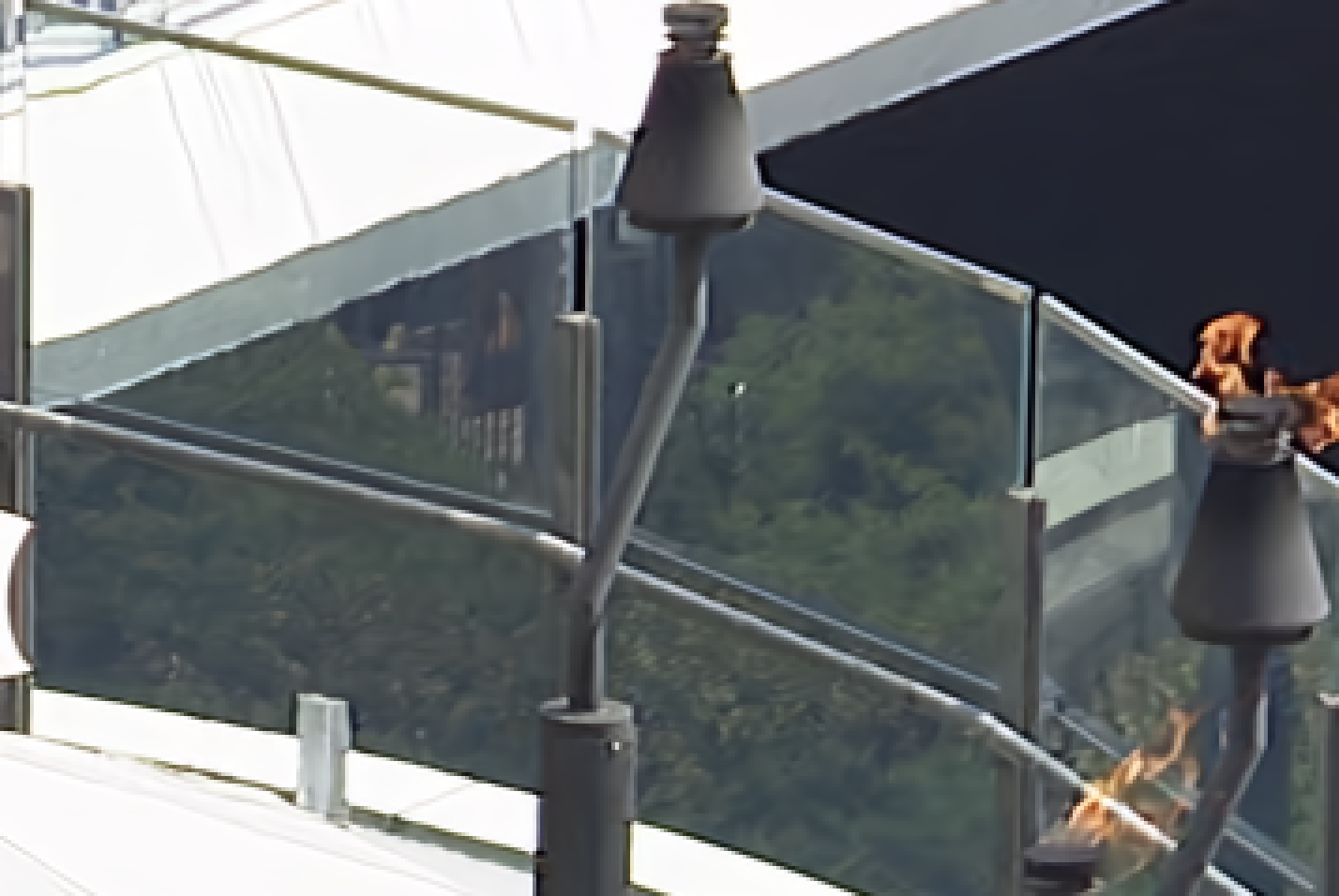}
        \vspace{-0.28em}
        \parbox[t][2.05em][t]{\linewidth}{\centering
            \scriptsize Ours\\[-0.18em]
            \scriptsize (bpp: 0.1651, psnr: 32.97)}
    \end{minipage}

    \vspace{-0.18em}
    {\small (b) HEVC Class B \textit{BQTerrace}}

    \vspace{0.18em}

    \begin{minipage}[t]{\imgw}
        \centering
        \includegraphics[
            width=\linewidth,
            height=\imgh,
            trim=0cm 0.2cm 7.0cm 0.2cm,
            clip
        ]{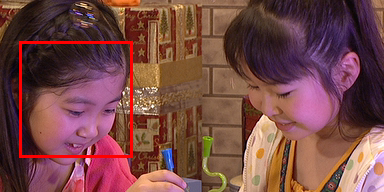}
        \vspace{-0.28em}
        \parbox[t][1.15em][t]{\linewidth}{\centering\scriptsize Original}
    \end{minipage}
    \begin{minipage}[t]{\imgw}
        \centering
        \includegraphics[width=\linewidth,height=\imgh]{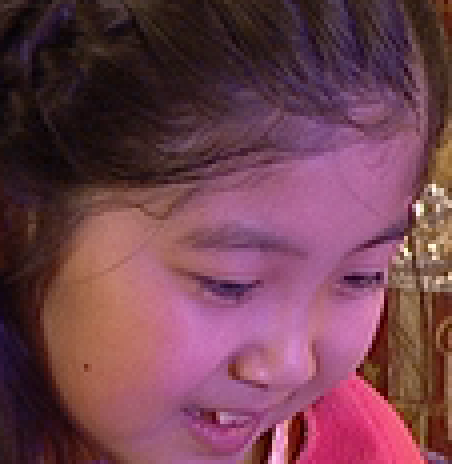}
        \vspace{-0.28em}
        \parbox[t][1.15em][t]{\linewidth}{\centering\scriptsize Ground Truth}
    \end{minipage}
    \begin{minipage}[t]{\imgw}
        \centering
        \includegraphics[width=\linewidth,height=\imgh]{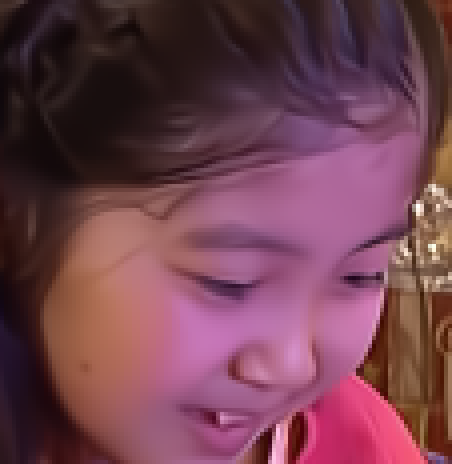}
        \vspace{-0.28em}
        \parbox[t][2.05em][t]{\linewidth}{\centering
            \scriptsize DCVC\\[-0.18em]
            \scriptsize (bpp: 0.2068, psnr: 31.43)}
    \end{minipage}
    \begin{minipage}[t]{\imgw}
        \centering
        \includegraphics[width=\linewidth,height=\imgh]{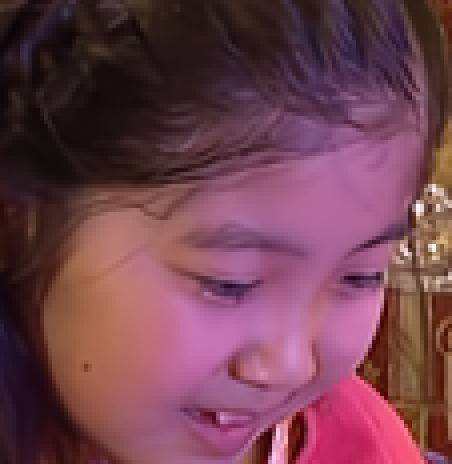}
        \vspace{-0.28em}
        \parbox[t][2.05em][t]{\linewidth}{\centering
            \scriptsize Ours\\[-0.18em]
            \scriptsize (bpp: 0.1998, psnr: 32.37)}
    \end{minipage}

    \vspace{-0.18em}
    {\small (c) HEVC Class D \textit{BlowingBubbles}}

    \vspace{-0.1em}
    \caption{Subjective quality comparison. From left to right: original frame with the selected region, cropped ground truth, baseline reconstruction, and reconstruction by the proposed method.}
    \label{fig:subjective_compare}
\end{figure}

Fig.~\ref{fig:minigop_alignment} shows the mini-GOP budget alignment result on a representative video from the HEVC Class B dataset, where the blue bars denote the total mini-GOP budget and the orange bars denote the actual expenditure. It can be observed that, after the base PI controller has already ensured that the overall budget remains under control, the proposed adjustment controller is able to keep the local expenditure close to the allocated budget while still flexibly redistributing bits within a short temporal window. This behavior helps explain why the proposed method can further improve the overall RD performance while preserving budget consistency.

Fig.~\ref{fig:subjective_compare} presents the subjective reconstruction quality comparison between the proposed full method and the original DCVC baseline on the DCVC backbone. For the same frame, the proposed method achieves higher PSNR at a lower bpp, while also showing clearly better recovery of texture details than the baseline. These results further verify that the proposed method can effectively improve subjective reconstruction quality and detail fidelity while maintaining budget consistency.

Meanwhile, the introduced DualGRU adjustment controller has low model complexity and inference overhead. The controller contains only 88.2K parameters. Measured on an NVIDIA RTX 4090 GPU, the average inference latency of a single forward step is only 0.357~ms. These results suggest that the proposed budget-constrained RD-optimized adjustment controller provides a favorable trade-off between rate--distortion gain and practical deployment cost.

\section{Conclusion}
\label{sec:conclusion}

We have proposed a feedback-driven rate control framework for learned video compression. By introducing the classical PI/PID control mechanism into the $\lambda$ domain, the proposed method enables frame-wise online bitrate regulation based on coding feedback, thereby establishing stable target bitrate tracking under a single-model multi-rate setting. On top of this, we further introduce a budget-constrained RD-optimized adjustment controller, which performs bounded residual refinement on the base control signal so as to optimize local inter-frame bit allocation while preserving overall budget consistency.

Experimental results demonstrate that the proposed method consistently achieves stable rate--distortion gains on multiple datasets while maintaining high bitrate tracking accuracy, verifying the effectiveness of combining explicit feedback regulation with budget-constrained optimization. The overall framework realizes unified variable-rate control and rate--distortion optimization without introducing additional decoding-side overhead.

In future work, we will further explore longer-term budget modeling and global bit allocation strategies across GOPs, and extend the framework to more complex coding structures and practical streaming scenarios, with the goal of achieving more refined sequence-level rate--distortion optimization.

\begin{acks}
This work was supported by the Postgraduate Innovative Project of Central South University.
\end{acks}

\bibliographystyle{ACM-Reference-Format}
\bibliography{myreference}

\end{document}